% VANILLA.STY
% COPYRIGHT (C) 1985, 1986 BY MICHAEL SPIVAK
% version date 1/1/86
\catcode`\@=11
\font\tensmc=cmcsc10      %change to CM fonts 3-31-87
%\font\tensmc=amcsc10
\def\smc{\tensmc}

\def\hcorrection#1{\advance\hoffset by #1 }
\def\vcorrection#1{\advance\voffset by #1 }
\def\wlog#1{}
\newif\iftitle@
\outer\def\title{\title@true\vglue 24\p@ plus 12\p@ minus 12\p@
   \bgroup\let\\=\cr\tabskip\centering
   \halign to \hsize\bgroup\tenbf\hfill\ignorespaces##\unskip\hfill\cr}
\def\endtitle{\cr\egroup\egroup\vglue 18\p@ plus 12\p@ minus 6\p@}
\outer\def\author{\iftitle@\vglue -18\p@ plus -12\p@ minus -6\p@\fi\vglue
    12\p@ plus 6\p@ minus 3\p@\bgroup\let\\=\cr\tabskip\centering
    \halign to \hsize\bgroup\smc\hfill\ignorespaces##\unskip\hfill\cr}
\def\endauthor{\cr\egroup\egroup\vglue 18\p@ plus 12\p@ minus 6\p@}
\outer\def\heading{\bigbreak\bgroup\let\\=\cr\tabskip\centering
    \halign to \hsize\bgroup\smc\hfill\ignorespaces##\unskip\hfill\cr}
\def\endheading{\cr\egroup\egroup\nobreak\medskip}
\outer\def\subheading#1{\medbreak\noindent{\tenbf\ignorespaces
      #1\unskip.\enspace}\ignorespaces}
\outer\def\proclaim#1{\medbreak\noindent\smc\ignorespaces
    #1\unskip.\enspace\sl\ignorespaces}
\outer\def\endproclaim{\par\ifdim\lastskip<\medskipamount\removelastskip
  \penalty 55 \fi\medskip\rm}
\outer\def\demo#1{\par\ifdim\lastskip<\smallskipamount\removelastskip
    \smallskip\fi\noindent{\smc\ignorespaces#1\unskip:\enspace}\rm
      \ignorespaces}
\outer\def\enddemo{\par\smallskip}
\newcount\footmarkcount@
\footmarkcount@=1
\def\makefootnote@#1#2{\insert\footins{\interlinepenalty=100
  \splittopskip=\ht\strutbox \splitmaxdepth=\dp\strutbox 
  \floatingpenalty=\@MM
  \leftskip=\z@\rightskip=\z@\spaceskip=\z@\xspaceskip=\z@
  \noindent{#1}\footstrut\rm\ignorespaces #2\strut}}
\def\footnote{\let\@sf=\empty\ifhmode\edef\@sf{\spacefactor
   =\the\spacefactor}\/\fi\futurelet\next\footnote@}
\def\footnote@{\ifx"\next\let\next\footnote@@\else
    \let\next\footnote@@@\fi\next}
\def\footnote@@"#1"#2{#1\@sf\relax\makefootnote@{#1}{#2}}
\def\footnote@@@#1{$^{\number\footmarkcount@}$\makefootnote@
   {$^{\number\footmarkcount@}$}{#1}\global\advance\footmarkcount@ by 1 }

\hyphenation{man-u-script man-u-scripts ap-pen-dix ap-pen-di-ces}
\hyphenation{data-base data-bases}
\ifx\amstexloaded@\relax\catcode`\@=13 
  \endinput\else\let\amstexloaded@=\relax\fi
\newlinechar=`\^^J
\def\eat@#1{}
\def\Space@.{\futurelet\Space@\relax}
\Space@. % 
\newhelp\athelp@
{Only certain combinations beginning with @ make sense to me.^^J
Perhaps you wanted \string\@\space for a printed @?^^J
I've ignored the character or group after @.}
\def\futureletnextat@{\futurelet\next\at@}
{\catcode`\@=\active
\lccode`\Z=`\@ \lowercase
{\gdef@{\expandafter\csname futureletnextatZ\endcsname}
\expandafter\gdef\csname atZ\endcsname
   {\ifcat\noexpand\next a\def\next{\csname atZZ\endcsname}\else
   \ifcat\noexpand\next0\def\next{\csname atZZ\endcsname}\else
    \def\next{\csname atZZZ\endcsname}\fi\fi\next}
\expandafter\gdef\csname atZZ\endcsname#1{\expandafter
   \ifx\csname #1Zat\endcsname\relax\def\next
     {\errhelp\expandafter=\csname athelpZ\endcsname
      \errmessage{Invalid use of \string@}}\else
       \def\next{\csname #1Zat\endcsname}\fi\next}
\expandafter\gdef\csname atZZZ\endcsname#1{\errhelp
    \expandafter=\csname athelpZ\endcsname
      \errmessage{Invalid use of \string@}}}}
\def\atdef@#1{\expandafter\def\csname #1@at\endcsname}
\newhelp\defahelp@{If you typed \string\define\space cs instead of
\string\define\string\cs\space^^J
I've substituted an inaccessible control sequence so that your^^J
definition will be completed without mixing me up too badly.^^J
If you typed \string\define{\string\cs} the inaccessible control sequence^^J
was defined to be \string\cs, and the rest of your^^J
definition appears as input.}
\newhelp\defbhelp@{I've ignored your definition, because it might^^J
conflict with other uses that are important to me.}
\def\define{\futurelet\next\define@}
\def\define@{\ifcat\noexpand\next\relax
  \def\next{\define@@}%
  \else\errhelp=\defahelp@
  \errmessage{\string\define\space must be followed by a control 
     sequence}\def\next{\def\garbage@}\fi\next}
\def\undefined@{}
\def\preloaded@{}    
\def\define@@#1{\ifx#1\relax\errhelp=\defbhelp@
   \errmessage{\string#1\space is already defined}\def\next{\def\garbage@}%
   \else\expandafter\ifx\csname\expandafter\eat@\string
         #1@\endcsname\undefined@\errhelp=\defbhelp@
   \errmessage{\string#1\space can't be defined}\def\next{\def\garbage@}%
   \else\expandafter\ifx\csname\expandafter\eat@\string#1\endcsname\relax
     \def\next{\def#1}\else\errhelp=\defbhelp@
     \errmessage{\string#1\space is already defined}\def\next{\def\garbage@}%
      \fi\fi\fi\next}
\def\famzero{\fam\z@}

\def\lim{\mathop{\famzero lim}}

\def\textfont@#1#2{\def#1{\relax\ifmmode
    \errmessage{Use \string#1\space only in text}\else#2\fi}}
\textfont@\rm\tenrm
\textfont@\it\tenit
\textfont@\sl\tensl
\textfont@\bf\tenbf
\textfont@\smc\tensmc
\let\ic@=\/
\def\/{\unskip\ic@}
\def\textfonti{\the\textfont1 }
\def\t#1#2{{\edef\next{\the\font}\textfonti\accent"7F \next#1#2}}
\let\B=\=
\let\D=\.
\def~{\unskip\nobreak\ \ignorespaces}
{\catcode`\@=\active
\gdef\@{\char'100 }}
\atdef@-{\leavevmode\futurelet\next\athyph@}
\def\athyph@{\ifx\next-\let\next=\athyph@@
  \else\let\next=\athyph@@@\fi\next}
\def\athyph@@@{\hbox{-}}
\def\athyph@@#1{\futurelet\next\athyph@@@@}
\def\athyph@@@@{\if\next-\def\next##1{\hbox{---}}\else
    \def\next{\hbox{--}}\fi\next}
\def\.{.\spacefactor=\@m}
\atdef@.{\null.}
\atdef@,{\null,}
\atdef@;{\null;}
\atdef@:{\null:}
\atdef@?{\null?}
\atdef@!{\null!}   
\def\srdr@{\thinspace}                     
\def\drsr@{\kern.02778em}
\def\sldl@{\kern.02778em}
\def\dlsl@{\thinspace}
\atdef@"{\unskip\futurelet\next\atqq@}
\def\atqq@{\ifx\next\Space@\def\next. {\atqq@@}\else
         \def\next.{\atqq@@}\fi\next.}
\def\atqq@@{\futurelet\next\atqq@@@}
\def\atqq@@@{\ifx\next`\def\next`{\atqql@}\else\def\next'{\atqqr@}\fi\next}
\def\atqql@{\futurelet\next\atqql@@}
\def\atqql@@{\ifx\next`\def\next`{\sldl@``}\else\def\next{\dlsl@`}\fi\next}
\def\atqqr@{\futurelet\next\atqqr@@}
\def\atqqr@@{\ifx\next'\def\next'{\srdr@''}\else\def\next{\drsr@'}\fi\next}

\def\textfontii{\the\textfont2 }
\def\{{\relax\ifmmode\lbrace\else
    {\textfontii f}\spacefactor=\@m\fi}
\def\}{\relax\ifmmode\rbrace\else
    \let\@sf=\empty\ifhmode\edef\@sf{\spacefactor=\the\spacefactor}\fi
      {\textfontii g}\@sf\relax\fi}   
\def\nonhmodeerr@#1{\errmessage
     {\string#1\space allowed only within text}}
\def\linebreak{\relax\ifhmode\unskip\break\else
    \nonhmodeerr@\linebreak\fi}
\def\allowlinebreak{\relax
   \ifhmode\allowbreak\else\nonhmodeerr@\allowlinebreak\fi}
\newskip\saveskip@
\def\nolinebreak{\relax\ifhmode\saveskip@=\lastskip\unskip
  \nobreak\ifdim\saveskip@>\z@\hskip\saveskip@\fi
   \else\nonhmodeerr@\nolinebreak\fi}
\def\newline{\relax\ifhmode\null\hfil\break
    \else\nonhmodeerr@\newline\fi}
\def\nonmathaerr@#1{\errmessage
     {\string#1\space is not allowed in display math mode}}
\def\nonmathberr@#1{\errmessage{\string#1\space is allowed only in math mode}}
\def\mathbreak{\relax\ifmmode\ifinner\break\else
   \nonmathaerr@\mathbreak\fi\else\nonmathberr@\mathbreak\fi}
\def\nomathbreak{\relax\ifmmode\ifinner\nobreak\else
    \nonmathaerr@\nomathbreak\fi\else\nonmathberr@\nomathbreak\fi}
\def\allowmathbreak{\relax\ifmmode\ifinner\allowbreak\else
     \nonmathaerr@\allowmathbreak\fi\else\nonmathberr@\allowmathbreak\fi}
\def\pagebreak{\relax\ifmmode
   \ifinner\errmessage{\string\pagebreak\space
     not allowed in non-display math mode}\else\postdisplaypenalty-\@M\fi
   \else\ifvmode\penalty-\@M\else\edef\spacefactor@
       {\spacefactor=\the\spacefactor}\vadjust{\penalty-\@M}\spacefactor@
        \relax\fi\fi}
\def\nopagebreak{\relax\ifmmode
     \ifinner\errmessage{\string\nopagebreak\space
    not allowed in non-display math mode}\else\postdisplaypenalty\@M\fi
    \else\ifvmode\nobreak\else\edef\spacefactor@
        {\spacefactor=\the\spacefactor}\vadjust{\penalty\@M}\spacefactor@
         \relax\fi\fi}
\def\newpage{\relax\ifvmode\vfill\penalty-\@M\else\nonvmodeerr@\newpage\fi}
\def\nonvmodeerr@#1{\errmessage
    {\string#1\space is allowed only between paragraphs}}
\def\smallpagebreak{\relax\ifvmode\smallbreak
      \else\nonvmodeerr@\smallpagebreak\fi}
\def\medpagebreak{\relax\ifvmode\medbreak
       \else\nonvmodeerr@\medpagebreak\fi}
\def\bigpagebreak{\relax\ifvmode\bigbreak
      \else\nonvmodeerr@\bigpagebreak\fi}
\newdimen\captionwidth@
\captionwidth@=\hsize
\advance\captionwidth@ by -1.5in
\def\caption#1{}
\def\topspace#1{\gdef\thespace@{#1}\ifvmode\def\next
    {\futurelet\next\topspace@}\else\def\next{\nonvmodeerr@\topspace}\fi\next}
\def\topspace@{\ifx\next\Space@\def\next. {\futurelet\next\topspace@@}\else
     \def\next.{\futurelet\next\topspace@@}\fi\next.}
\def\topspace@@{\ifx\next\caption\let\next\topspace@@@\else
    \let\next\topspace@@@@\fi\next}
 \def\topspace@@@@{\topinsert\vbox to 
       \thespace@{}\endinsert}
\def\topspace@@@\caption#1{\topinsert\vbox to
    \thespace@{}\nobreak
      \smallskip
    \setbox\z@=\hbox{\noindent\ignorespaces#1\unskip}%
   \ifdim\wd\z@>\captionwidth@
   \centerline{\vbox{\hsize=\captionwidth@\noindent\ignorespaces#1\unskip}}%
   \else\centerline{\box\z@}\fi\endinsert}
\def\midspace#1{\gdef\thespace@{#1}\ifvmode\def\next
    {\futurelet\next\midspace@}\else\def\next{\nonvmodeerr@\midspace}\fi\next}
\def\midspace@{\ifx\next\Space@\def\next. {\futurelet\next\midspace@@}\else
     \def\next.{\futurelet\next\midspace@@}\fi\next.}
\def\midspace@@{\ifx\next\caption\let\next\midspace@@@\else
    \let\next\midspace@@@@\fi\next}
 \def\midspace@@@@{\midinsert\vbox to 
       \thespace@{}\endinsert}
\def\midspace@@@\caption#1{\midinsert\vbox to
    \thespace@{}\nobreak
      \smallskip
      \setbox\z@=\hbox{\noindent\ignorespaces#1\unskip}%
      \ifdim\wd\z@>\captionwidth@
    \centerline{\vbox{\hsize=\captionwidth@\noindent\ignorespaces#1\unskip}}%
    \else\centerline{\box\z@}\fi\endinsert}
\mathchardef\prime@="0230
\def\prime{{{}\prime@{}}}
\def\prim@s{\prime@\futurelet\next\pr@m@s}

\def\,{\relax\ifmmode\mskip\thinmuskip\else\thinspace\fi}
\def\!{\relax\ifmmode\mskip-\thinmuskip\else\negthinspace\fi}
\def\frac#1#2{{#1\over#2}}

\def\:{\nobreak\hskip.1111em{:}\hskip.3333em plus .0555em\relax}
\def\intic@{\mathchoice{\hskip5\p@}{\hskip4\p@}{\hskip4\p@}{\hskip4\p@}}
\def\negintic@
 {\mathchoice{\hskip-5\p@}{\hskip-4\p@}{\hskip-4\p@}{\hskip-4\p@}}
\def\intkern@{\mathchoice{\!\!\!}{\!\!}{\!\!}{\!\!}}
\def\intdots@{\mathchoice{\cdots}{{\cdotp}\mkern1.5mu
    {\cdotp}\mkern1.5mu{\cdotp}}{{\cdotp}\mkern1mu{\cdotp}\mkern1mu
      {\cdotp}}{{\cdotp}\mkern1mu{\cdotp}\mkern1mu{\cdotp}}}
\newcount\intno@             
\def\iint{\intno@=\tw@\futurelet\next\ints@} 
\def\iiint{\intno@=\thr@@\futurelet\next\ints@}
\def\iiiint{\intno@=4 \futurelet\next\ints@}
\def\idotsint{\intno@=\z@\futurelet\next\ints@}
\def\ints@{\findlimits@\ints@@}
\newif\iflimtoken@
\newif\iflimits@
\def\findlimits@{\limtoken@false\limits@false\ifx\next\limits
 \limtoken@true\limits@true\else\ifx\next\nolimits\limtoken@true\limits@false
    \fi\fi}
\def\multintlimits@{\intop\ifnum\intno@=\z@\intdots@
  \else\intkern@\fi
    \ifnum\intno@>\tw@\intop\intkern@\fi
     \ifnum\intno@>\thr@@\intop\intkern@\fi\intop}
\def\multint@{\int\ifnum\intno@=\z@\intdots@\else\intkern@\fi
   \ifnum\intno@>\tw@\int\intkern@\fi
    \ifnum\intno@>\thr@@\int\intkern@\fi\int}
\def\ints@@{\iflimtoken@\def\ints@@@{\iflimits@
   \negintic@\mathop{\intic@\multintlimits@}\limits\else
    \multint@\nolimits\fi\eat@}\else
     \def\ints@@@{\multint@\nolimits}\fi\ints@@@}
\def\Sb{_\bgroup\vspace@
        \baselineskip=\fontdimen10 \scriptfont\tw@
        \advance\baselineskip by \fontdimen12 \scriptfont\tw@
        \lineskip=\thr@@\fontdimen8 \scriptfont\thr@@
        \lineskiplimit=\thr@@\fontdimen8 \scriptfont\thr@@
        \Let@\vbox\bgroup\halign\bgroup \hfil$\scriptstyle
            {##}$\hfil\cr}
\def\endSb{\crcr\egroup\egroup\egroup}
\def\Sp{^\bgroup\vspace@
        \baselineskip=\fontdimen10 \scriptfont\tw@
        \advance\baselineskip by \fontdimen12 \scriptfont\tw@
        \lineskip=\thr@@\fontdimen8 \scriptfont\thr@@
        \lineskiplimit=\thr@@\fontdimen8 \scriptfont\thr@@
        \Let@\vbox\bgroup\halign\bgroup \hfil$\scriptstyle
            {##}$\hfil\cr}
\def\endSp{\crcr\egroup\egroup\egroup}
\def\Let@{\relax\iffalse{\fi\let\\=\cr\iffalse}\fi}
\def\vspace@{\def\vspace##1{\noalign{\vskip##1 }}}
\def\aligned{\,\vcenter\bgroup\vspace@\Let@\openup\jot\m@th\ialign
  \bgroup \strut\hfil$\displaystyle{##}$&$\displaystyle{{}##}$\hfil\crcr}
\def\endaligned{\crcr\egroup\egroup}
\def\matrix{\,\vcenter\bgroup\Let@\vspace@
    \normalbaselines
  \m@th\ialign\bgroup\hfil$##$\hfil&&\quad\hfil$##$\hfil\crcr
    \mathstrut\crcr\noalign{\kern-\baselineskip}}
\def\endmatrix{\crcr\mathstrut\crcr\noalign{\kern-\baselineskip}\egroup
                \egroup\,}
\newtoks\hashtoks@
\hashtoks@={#}
\def\format{\crcr\egroup\iffalse{\fi\ifnum`}=0 \fi\format@}
\def\format@#1\\{\def\preamble@{#1}%
  \def\c{\hfil$\the\hashtoks@$\hfil}%
  \def\r{\hfil$\the\hashtoks@$}%
  \def\l{$\the\hashtoks@$\hfil}%
  \setbox\z@=\hbox{\xdef\Preamble@{\preamble@}}\ifnum`{=0 \fi\iffalse}\fi
   \ialign\bgroup\span\Preamble@\crcr}

\def\cases{\left\{\,\vcenter\bgroup\vspace@
     \normalbaselines\openup\jot\m@th
       \Let@\ialign\bgroup$##$\hfil&\quad$##$\hfil\crcr
      \mathstrut\crcr\noalign{\kern-\baselineskip}}

\newif\iftagsleft@
\tagsleft@true
\def\TagsOnRight{\global\tagsleft@false}
\def\tag#1$${\iftagsleft@\leqno\else\eqno\fi
 \hbox{\def\pagebreak{\global\postdisplaypenalty-\@M}%
 \def\nopagebreak{\global\postdisplaypenalty\@M}\rm(#1\unskip)}%
  $$\postdisplaypenalty\z@\ignorespaces}
\interdisplaylinepenalty=\@M
\def\allowdisplaybreak@{\def\allowdisplaybreak{\noalign{\allowbreak}}}
\def\displaybreak@{\def\displaybreak{\noalign{\break}}}
\def\align#1\endalign{\def\tag{&}\vspace@\allowdisplaybreak@\displaybreak@
  \iftagsleft@\lalign@#1\endalign\else
   \ralign@#1\endalign\fi}
\def\ralign@#1\endalign{\displ@y\Let@\tabskip\centering\halign to\displaywidth
     {\hfil$\displaystyle{##}$\tabskip=\z@&$\displaystyle{{}##}$\hfil
       \tabskip=\centering&\llap{\hbox{(\rm##\unskip)}}\tabskip\z@\crcr
             #1\crcr}}
\def\lalign@
 #1\endalign{\displ@y\Let@\tabskip\centering\halign to \displaywidth
   {\hfil$\displaystyle{##}$\tabskip=\z@&$\displaystyle{{}##}$\hfil
   \tabskip=\centering&\kern-\displaywidth
        \rlap{\hbox{(\rm##\unskip)}}\tabskip=\displaywidth\crcr
               #1\crcr}}
\def\overrightarrow{\mathpalette\overrightarrow@}
\def\overrightarrow@#1#2{\vbox{\ialign{$##$\cr
    #1{-}\mkern-6mu\cleaders\hbox{$#1\mkern-2mu{-}\mkern-2mu$}\hfill
     \mkern-6mu{\to}\cr
     \noalign{\kern -1\p@\nointerlineskip}
     \hfil#1#2\hfil\cr}}}
\def\overleftarrow{\mathpalette\overleftarrow@}
\def\overleftarrow@#1#2{\vbox{\ialign{$##$\cr
     #1{\leftarrow}\mkern-6mu\cleaders\hbox{$#1\mkern-2mu{-}\mkern-2mu$}\hfill
      \mkern-6mu{-}\cr
     \noalign{\kern -1\p@\nointerlineskip}
     \hfil#1#2\hfil\cr}}}
\def\overleftrightarrow{\mathpalette\overleftrightarrow@}
\def\overleftrightarrow@#1#2{\vbox{\ialign{$##$\cr
     #1{\leftarrow}\mkern-6mu\cleaders\hbox{$#1\mkern-2mu{-}\mkern-2mu$}\hfill
       \mkern-6mu{\to}\cr
    \noalign{\kern -1\p@\nointerlineskip}
      \hfil#1#2\hfil\cr}}}
\def\underrightarrow{\mathpalette\underrightarrow@}
\def\underrightarrow@#1#2{\vtop{\ialign{$##$\cr
    \hfil#1#2\hfil\cr
     \noalign{\kern -1\p@\nointerlineskip}
    #1{-}\mkern-6mu\cleaders\hbox{$#1\mkern-2mu{-}\mkern-2mu$}\hfill
     \mkern-6mu{\to}\cr}}}
\def\underleftarrow{\mathpalette\underleftarrow@}
\def\underleftarrow@#1#2{\vtop{\ialign{$##$\cr
     \hfil#1#2\hfil\cr
     \noalign{\kern -1\p@\nointerlineskip}
     #1{\leftarrow}\mkern-6mu\cleaders\hbox{$#1\mkern-2mu{-}\mkern-2mu$}\hfill
      \mkern-6mu{-}\cr}}}
\def\underleftrightarrow{\mathpalette\underleftrightarrow@}
\def\underleftrightarrow@#1#2{\vtop{\ialign{$##$\cr
      \hfil#1#2\hfil\cr
    \noalign{\kern -1\p@\nointerlineskip}
     #1{\leftarrow}\mkern-6mu\cleaders\hbox{$#1\mkern-2mu{-}\mkern-2mu$}\hfill
       \mkern-6mu{\to}\cr}}}
\def\sqrt#1{\radical"270370 {#1}}
\def\dots{\relax\ifmmode\let\next=\ldots\else\let\next=\tdots@\fi\next}
\def\tdots@{\unskip\ \tdots@@}
\def\tdots@@{\futurelet\next\tdots@@@}
\def\tdots@@@{$\mathinner{\ldotp\ldotp\ldotp}\,
   \ifx\next,$\else
   \ifx\next.\,$\else
   \ifx\next;\,$\else
   \ifx\next:\,$\else
   \ifx\next?\,$\else
   \ifx\next!\,$\else
   $ \fi\fi\fi\fi\fi\fi}
\def\text{\relax\ifmmode\let\next=\text@\else\let\next=\text@@\fi\next}
\def\text@@#1{\hbox{#1}}
\def\text@#1{\mathchoice
 {\hbox{\everymath{\displaystyle}\def\textfonti{\the\textfont1 }%
    \def\textfontii{\the\textfont2 }\textdef@@ T#1}}
 {\hbox{\everymath{\textstyle}\def\textfonti{\the\textfont1 }%
    \def\textfontii{\the\textfont2 }\textdef@@ T#1}}
 {\hbox{\everymath{\scriptstyle}\def\textfonti{\the\scriptfont1 }%
   \def\textfontii{\the\scriptfont2 }\textdef@@ S\rm#1}}
 {\hbox{\everymath{\scriptscriptstyle}\def\textfonti{\the\scriptscriptfont1 }%
   \def\textfontii{\the\scriptscriptfont2 }\textdef@@ s\rm#1}}}
\def\textdef@@#1{\textdef@#1\rm \textdef@#1\bf
   \textdef@#1\sl \textdef@#1\it}

\def\textdef@#1#2{\def\next{\csname\expandafter\eat@\string#2fam\endcsname}%
\if S#1\edef#2{\the\scriptfont\next\relax}%
 \else\if s#1\edef#2{\the\scriptscriptfont\next\relax}%
 \else\edef#2{\the\textfont\next\relax}\fi\fi}
\scriptfont\itfam=\tenit \scriptscriptfont\itfam=\tenit
\scriptfont\slfam=\tensl \scriptscriptfont\slfam=\tensl
\mathcode`\0="0030
\mathcode`\1="0031
\mathcode`\2="0032
\mathcode`\3="0033
\mathcode`\4="0034
\mathcode`\5="0035
\mathcode`\6="0036
\mathcode`\7="0037
\mathcode`\8="0038
\mathcode`\9="0039
\def\Cal{\relax\ifmmode\let\next=\Cal@\else
     \def\next{\errmessage{Use \string\Cal\space only in math mode}}\fi\next}
\def\Cal@#1{{\fam2 #1}}
\def\bold{\relax\ifmmode\let\next=\bold@\else
   \def\next{\errmessage{Use \string\bold\space only in math
      mode}}\fi\next}\def\bold@#1{{\fam\bffam #1}}
\mathchardef\Gamma="0000
\mathchardef\Delta="0001
\mathchardef\Theta="0002
\mathchardef\Lambda="0003
\mathchardef\Xi="0004
\mathchardef\Pi="0005
\mathchardef\Sigma="0006
\mathchardef\Upsilon="0007
\mathchardef\Phi="0008
\mathchardef\Psi="0009
\mathchardef\Omega="000A
\mathchardef\varGamma="0100
\mathchardef\varDelta="0101
\mathchardef\varTheta="0102
\mathchardef\varLambda="0103
\mathchardef\varXi="0104
\mathchardef\varPi="0105
\mathchardef\varSigma="0106
\mathchardef\varUpsilon="0107
\mathchardef\varPhi="0108
\mathchardef\varPsi="0109
\mathchardef\varOmega="010A
\font\dummyft@=dummy
\fontdimen1 \dummyft@=\z@
\fontdimen2 \dummyft@=\z@
\fontdimen3 \dummyft@=\z@
\fontdimen4 \dummyft@=\z@
\fontdimen5 \dummyft@=\z@
\fontdimen6 \dummyft@=\z@
\fontdimen7 \dummyft@=\z@
\fontdimen8 \dummyft@=\z@
\fontdimen9 \dummyft@=\z@
\fontdimen10 \dummyft@=\z@
\fontdimen11 \dummyft@=\z@
\fontdimen12 \dummyft@=\z@
\fontdimen13 \dummyft@=\z@
\fontdimen14 \dummyft@=\z@
\fontdimen15 \dummyft@=\z@
\fontdimen16 \dummyft@=\z@
\fontdimen17 \dummyft@=\z@
\fontdimen18 \dummyft@=\z@
\fontdimen19 \dummyft@=\z@
\fontdimen20 \dummyft@=\z@
\fontdimen21 \dummyft@=\z@
\fontdimen22 \dummyft@=\z@
\def\fontlist@{\\{\tenrm}\\{\sevenrm}\\{\fiverm}\\{\teni}\\{\seveni}%
 \\{\fivei}\\{\tensy}\\{\sevensy}\\{\fivesy}\\{\tenex}\\{\tenbf}\\{\sevenbf}%
 \\{\fivebf}\\{\tensl}\\{\tenit}\\{\tensmc}}
\def\dodummy@{{\def\\##1{\global\let##1=\dummyft@}\fontlist@}}
\newif\ifsyntax@
\newcount\countxviii@
\def\newtoks@{\alloc@5\toks\toksdef\@cclvi}
\def\nopages@{\output={\setbox\z@=\box\@cclv \deadcycles=\z@}\newtoks@\output}
\def\syntax{\syntax@true\dodummy@\countxviii@=\count18
\loop \ifnum\countxviii@ > \z@ \textfont\countxviii@=\dummyft@
   \scriptfont\countxviii@=\dummyft@ \scriptscriptfont\countxviii@=\dummyft@
     \advance\countxviii@ by-\@ne\repeat
\dummyft@\tracinglostchars=\z@
  \nopages@\frenchspacing\hbadness=\@M}
\def\magstep#1{\ifcase#1 1000\or
 1200\or 1440\or 1728\or 2074\or 2488\or 
 \errmessage{\string\magstep\space only works up to 5}\fi\relax}
{\lccode`\2=`\p \lccode`\3=`\t 
 \lowercase{\gdef\tru@#123{#1truept}}}

\def\scaletype#1{\mag=#1\relax
 \hsize=\expandafter\tru@\the\hsize
 \vsize=\expandafter\tru@\the\vsize
 \dimen\footins=\expandafter\tru@\the\dimen\footins}

\def\scalefont#1#2\andcallit#3{\edef\font@{\the\font}#1\font#3=
  \fontname\font\space scaled #2\relax\font@}
\def\Mag@#1#2{\ifdim#1<1pt\multiply#1 #2\relax\divide#1 1000 \else
  \ifdim#1<10pt\divide#1 10 \multiply#1 #2\relax\divide#1 100\else
  \divide#1 100 \multiply#1 #2\relax\divide#1 10 \fi\fi}
\def\scalelinespacing#1{\Mag@\baselineskip{#1}\Mag@\lineskip{#1}%
  \Mag@\lineskiplimit{#1}}
\def\wlog#1{\immediate\write-1{#1}}
\catcode`\@=\active

%Springer format:
% \input lecmono.cmm   % these are the macros for Lect. Notes in Phys
\magnification 1200
\hsize=15.2truecm
%\advance\voffset by 1truecm
%\advance\hoffset by .2truecm
%\parskip 0pt plus 1pt
\overfullrule=0pt
%\advance\baselineskip by 1pt

%paper format:
%\magnification 1200 %optional
%\hsize=16.6truecm%15.6truecm
\vsize=22.6truecm %%%cancel optionally

\font\eightrm=cmr8

\font\gross=cmcsc10     
\font\gr=cmcsc10

%\font\twelvebf=cmbx12    
%\font\twbf=cmb10 scaled \magstep 1

 2      
\font\Titel=cmr10 scaled \magstep 2

%\nopagenumbers
%\footline={\hss}

%\TagsOnRight

% running head
%\csname lrh\endcsname
%\csname rrh\endcsname
%\newcount\titel
%\headline={\ifnum\titel=0\eightrm\ifodd\pageno\hss
%\rrh\qquad\ \folio\else\folio
%            \qquad\ \lrh
%            \hss\fi\else%
%             \hss\global\titel=0\fi}

%\headline{\centerline{\eightrm OQP -- \the\day/\the\month/\the\year\ --
%PAGE \folio}}
%\headline={\hss\tenrm\folio}      %\footline={\hss}
%\headline={\ifnum\titel=0\eightrm\ifodd\pageno\hss
%            ...\qquad\ \folio\else\folio
%            \qquad\ ...  
%            \hss\fi\else%
%             \hss\global\titel=0\fi}

\hyphenation{
Meas-ure-ment Meas-uring 
meas-ure-ment meas-ure meas-uring
pre-meas-ure-ment pre-meas-ure pre-meas-uring}

%long phrases
\def\sy{system}                       
                      
\def\qm{quantum mechanics}
\def\ty{theory}                       
\def\me{meas\-ure}                    \def\mt{\me{}ment}

\def\pov{{\gross pov} meas\-ure}      \def\pv{{\gross pv} \me}

\def\ob{observable}                  \def\obs{observables}

%operators:
\def\pee#1{{P[#1]}}           %projector on state
\def\bo#1{{\bold #1}}                    %for use in math mode

             %norm   
\def\ip#1#2{\left\langle\,#1\,|\,#2\,\right\rangle} %inner product
    %Dirac operators
                %Dirac ket
                %Dirac bra
\def\tr#1{\text{tr}\bigl[#1\bigr]}       %trace
           
    \def\ts12{{\textstyle{\frac 12}}}

%special letters
\def\Rea{{\bold R}}

\def\fii{\varphi}   
                       \def\cs{$\Cal S$}

%math objects
\def\hi{{\Cal H}}                     
\def\lh{\Cal L(\Cal H)}               \def\th{\Cal T(\Cal H)} 
                                      \def\sh{\Cal S(\Cal H)}
            
              \def\ha{\Cal H_{\Cal A}}
               \def\lha{\Cal L(\ha )}

\def\of{(\Omega ,\Cal F)}       \def\rb{\bigl(\Rea ,\Cal B(\Rea )\bigr)}

\def\pa{\Cal R_{\Cal S}}         \def\pas{\Cal R_{\Cal A}} %partial trace

\def\s{$\Cal S$}         \def\a{$\Cal A$}      \def\sa{$\Cal S + \Cal A$}

%\define\em{\langle\ha ,\allowmathbreak P_{\Cal A},
%\allowmathbreak T_{\Cal A},V,f\rangle}

\def\em{\langle\ha ,\allowmathbreak Z,
\allowmathbreak T_{\Cal A},V,f\rangle}

\define\m{$\Cal M$}
%\define\im{\Cal I_{\Cal M}}

\def\us{U(\varphi\otimes\phi )}      \def\vs{V(T\otimes T_{Cal A})}

\def\vas{\langle}                    \def\oik{\rangle}

                 \def\12pi{\frac 1{2\pi}}

\nopagenumbers

\advance\voffset by -1truecm
%\advance\baselineskip by 6pt

\parskip 0pt plus 1pt
%\advance\vsize by -0,9 true cm\advance\voffset by 0,9 true cm
%\advance\hsize by -1,26 true cm %\advance\hoffset by 1,26 true cm

\font\gross=cmcsc10
\font\Titel=cmb10 scaled \magstep 2
 1
\font\eightrm=cmr8

\def\R{\hbox{I\kern -0.2em R}}
\def\1{\hbox{1\kern -0.3em I}}

\headline{\noindent{\eightrm Correlation Properties of
Quantum Measurements -- 
Busch and Lahti %-- \the\month/\the\year\
\hfill \folio}}

\topinsert
\noindent{\eightrm J.\ Math.\ Phys. -- in print, March 1996}
\endinsert

\title{\Titel Correlation Properties of Quantum Measurements}
\endtitle

\vskip .7truecm
\author 
Paul Busch\footnote{
Department of Applied Mathematics, 
The University of Hull, Hull HU6 7RX, UK.\newline
{}\ E-mail: p.busch\@maths.hull.ac.uk}
and
Pekka J.\ Lahti\footnote{
Department of Physics, University of Turku, 20014 Turku,
Finland.\newline
{}\ E-mail: pekka.lahti\@utu.fi}
\endauthor

\vskip 1truecm
\centerline{\gross Abstract}
\vskip 3pt
\noindent
The kind of information provided by a \mt\ is determined
in terms of the correlation established between observables
of the apparatus and the measured system. Using the framework
of quantum \mt\ theory, necessary and sufficient conditions
for a \mt\ interaction to produce strong correlations are
given and are found to be related to properties of the final
object and apparatus states. These general results are illustrated
with reference to the {\it standard model} of the quantum theory
of \mt.

\vskip 5pt

%\footnote"{}"{}
%\footnote"{}"{
%\centerline{{\bf Turku-FL-P20-95}}
%}

\vfil\eject

%\bye

\subheading{I. Introduction}

\noindent
Any physical \mt\ is carried out in order to provide
information about a specified system, its state prior to or
after the \mt. The procedure generally is to ascertain the
values of a pointer \ob, which have become correlated
with some observable of the \me{}d \sy. Thus the kind of
information available by a given \mt\ depends on the
statistical dependencies established by the interaction
between the apparatus, or some probe system, and the object.

The minimal content of the notion of \mt\ in \qm\ is given
by the {\it probability reproducibility} requirement;
according to this condition, a particular \mt\ scheme qualifies as
a \mt\ of a given \ob\ $E$ if for all initial states of the object
system the associated probability distributions of $E$ are
reproduced in the resulting statistics of pointer readings$.^1$ 
Regarding a large ensemble of object plus apparatus
systems as one individual system, this situation can be
described in terms of strong correlations between the values
of the frequency operators [see, e.g., ref. 1] associated
with the observable $E$ and the pointer \ob, respectively.
In the present context we shall not be concerned with the
ensembles regarded as individual objects but rather we shall
analyze statistical dependencies between individual members of the
ensembles as they show up in certain correlation
quantities. The following three kinds of correlations
appear naturally in the \mt\ context: correlations between
an object \ob\ $E$ and the pointer observable; correlations
between the values of these \ob{}s; and correlations between
the conditional final states of the object system and apparatus,
respectively. Our goal is to give exhaustive
characterizations of the conditions under which these correlations
are established. It will be found that the final component
states of the object and apparatus must possess certain
properties in order that such correlations may be strong.

Our investigation builds on previous work published in ref. 2. 
Correcting an erroneous characterization of
strong correlations used in that paper, we give here a
complete account of necessary and sufficient conditions
for the occurrence of strong correlations. In addition, the
scope of the results is extended beyond unitary \mt{}s and
sharp \ob{}s to cover arbitrary \mt{}s and pointer \obs\ and
general object \obs. Finally, possible fields of
applications are indicated on the basis of the standard
model of \mt\ \ty, which was recently used in various
proposals for quantum and atomic optics {\gr QND} \mt{}s$.^3$

\subheading{II. Framework}

\subheading{2.1}
We follow here the Hilbert space formulation of quantum mechanics
in which the description of a physical system \s\ is based on a
complex separable Hilbert space $\hi$, with the inner product
$\ip{\cdot}{\cdot}$,
and which builds on the dual concepts of states and observables
reflecting the general structure of an experiment: preparation of the system
followed by a measurement.

\subheading{2.2}
Let $\lh$ denote the set of bounded linear operators on $\hi$  and  $\th$
its subset of trace class operators.
A {\it state}  of \s\ is given as a positive linear operator
$T:\hi\to\hi$  of trace one. We let 
$\sh := \{T\in\th\,|\, T\geq O,\ \tr{T}=1\}$ 
denote the set of states of \s,  and we recall that
$\sh$ is a ($\sigma$-)convex subset of $\th$, the one-dimensional
projection operators $\pee\fii$
(genereted by the unit vectors $\fii\in\hi$) being its
extremal elements. The $\pee\fii$ 
shall be called {\it vector states}.

\subheading{2.3}
Let $\Omega$ be a set and $\Cal F$ a $\sigma$-algebra of subsets of $\Omega$.
An {\it observable} of \s\ is represented as (and identified with)
a normalised positive operator valued measure, a \pov,
$E:\Cal F \to\lh$, that is, an operator valued mapping $X\mapsto E(X)$
on $\Cal F$ with the properties: $i)\ E(\Omega) = I$,
$ii)\ E(X)\geq O$, and $E(\cup X_i) = \sum E(X_i)$ for any disjoint
sequence $(X_i)\subset\Cal F$, with the series $\sum E(X_i)$ 
converging in the weak operator topology of $\lh$. 
We recall that the projection valued measures, the \pv{}s, are a particular case
of the \pov{}s; furthermore a \pov\ $E$ is a \pv, that is, $E(X)^2 = E(X)$ for all
$X\in\Cal F$, if and only if $E$ is multiplicative, that is,
$E(X\cap Y) = E(X)E(Y)$ for all $X,Y\in\Cal F$. 
Observables which are represented by \pv{} are  called {\it sharp} observables.
It is by now well-established  that the extension of
the notion of observables from  \pv{}s to  \pov{}s is a necessity
in \qm. 

\subheading{2.4}
The probability measure
$$
p^E_T:\Cal F\to [0,1],\ X \mapsto p^E_T(X) := \tr{TE(X)}\tag 1
$$
defined by an observable $E$ and a state $T$ 
is related to a measurement by virtue of the minimal interpretation
of quantum mechanics: {\it the number} $p^E_T(X)$ 
{\it is the probability that a measuremement of the observable E 
performed on the system \s\ in the state T leads to a result in the set X}.
The intended empirical content of  this statement is the following:
if the same $E$-measurement were repeated sufficiently many times under the
same conditions (characterised by $T$), then in the long run the relative
frequency of the occurrence of the measurement results in $X$ would approach
the number $p^E_T(X)$.

\subheading{III. Measurement}

\subheading{A. General}

\subheading{3.1}
A {\it measurement scheme}
for  the (object) system \s\ consists of 
a {\it measuring apparatus} \a\ [with its Hilbert space $\ha$], 
a {\it pointer observable}  $Z:\Cal F_{\Cal A}\to\lha$ 
[with its `space of values' $(\Omega_{\Cal A},\Cal F_{\Cal A})$],
an {\it initial state} $T_{\Cal A}$ of the apparatus,
a {\it meas\-urement coupling} $V$
[a linear state transformation
on  $\Cal T(\hi\otimes\ha)$], and 
a [meas\-urable] {\it pointer function} $f:\Omega_{\Cal A}\to\Omega$,
with the assumption that if $T\in\sh$ is an initial state of \s,
then $V(T\otimes T_{\Cal A})$ is the final state of the compound
object-apparatus system \sa.
Taking the partial traces of $\vs$ over $\ha$ and $\hi$, respectively,
one gets
the corresponding reduced states  $\pa(\vs)$ and  $\pas(\vs)$
of \s\ and \a, respectively; then
the probability measure of the pointer observable $Z$ in the final
apparatus state is completely determined as
$Y\mapsto p_{\pas(\vs)}^Z(Y)=\tr{\pas(\vs)Z(Y)}$, $Y\in\Cal F_{\Cal A}$.

\subheading{3.2}
A measurement scheme $\Cal M:=\em$  defines
an observable $E^{\Cal M}$ of \s\ with the space of values $(\Omega,\Cal F)$
via the relation:
for any $X\in\Cal F$ and $T\in\sh$,
$$
p^{E^{\Cal M}}_T(X) := p^Z_{\pas(\vs)}(f^{-1}(X)).\tag 2
$$
This is  the observable {\it measured} by means of the scheme
$\Cal M$ in the sense that the totality of the distributions
$p^Z_{\pas(\vs)}$ (for all $T\in\sh$) of the pointer outcomes
in the final apparatus states
determine the \pov\ $E^{\Cal M}$ via (2).
A  measurement scheme $\Cal M$
is a measurement of a given observable $E$ if 
the measured observable $E^{\Cal M}$ equals $E$.

\subheading{3.3}
There is an important subclass of measurement schemes for \s.
They consist of a sharp pointer observable $Z$, a vector state preparation
of \a, $T_{\Cal A} =\pee{\phi}$, $\phi\in\ha$, $\ip{\phi}{\phi} = 1$,
and a unitary measurement coupling $\vs = UT\otimes T_{\Cal A}U^*$,
with a unitary map $U$ on $\hi\otimes\ha$. Subsuming the possible pointer function
in the definition of $Z$ by identification of $\Omega$ and $\Omega_{\Cal
A}$, we denote such a scheme $\Cal M_U
:= \vas\ha,Z,\pee{\phi},U\oik$
and call it a {\it unitary measurement scheme} (with the understanding that
$Z$ is a sharp observable). It is a basic result of the quantum theory of
measurement that every observable $E$ of \s\ admits a unitary measurement,
that is, there is a unitary measurement scheme $\Cal M_U$ such that
$E^{\Cal M_U} = E.^4$ Thus the relation between \mt\ schemes and
\pov{}s induced by (2) defines a map from the former onto the latter.
In spite of this fundamental result, it is
important to consider general meaurement schemes $\Cal M$, since in many
applications the choices of a sharp pointer and a vector state preparation of
the apparatus are not physically realizable.

\subheading{3.4}
Another basic condition for a measurement scheme
$\em$ to qualify as a measurement
is the
requirement that the measurement should lead to a definite result.
We take this
requirement to entail, first of all, that the pointer
observable $Z$ should have assumed a definite value after the measurement.
One should then be able to `read the actual value' of the pointer
observable $Z$ and deduce from this the value of the measured 
observable. As well known, the task of explaining the occurrence of a definite
pointer value at the end of a measuring process presents one of the major open 
problems in quantum mechanics. We do not enter this
difficult question here.
(For an overview of the issues involved,
the reader may wish to consult ref. 1).
There are, however, some necessary conditions for the pointer observable
$Z$ to assume a definite value in the final apparatus state $\pas(\vs)$,
conditions which are tractable and which call for the study of the
correlation properties of a measurement.
These conditions are the subject of the present paper.

\subheading{B. Reading of pointer values}

\subheading{3.5}
The reading of \mt\ outcomes involves the discrimination between the
elements of a finite (or, as an idealization, countable) set
of alternative pointer values. In order to formulate this
idea in the  general context of an $E$-meas\-urement
$\Cal M$, we introduce the notion of a {\it reading scale}
as a countable partition of the value space of the pointer
observable, $\Omega_{\Cal A}$ = $\cup f^{-1}(X_i)$, induced
by a countable partition of the value space of the
meas\-ured observable, $\Omega$ = $\cup X_i$, $X_i\in\Cal
F$, $X_i\cap X_j = \emptyset$ for $i\ne j$. Such a reading
scale will be denoted $\Cal R$, and we let $\text{\bf I\/}$
denote its index set.

\subheading{3.6}
A reading scale $\Cal R$ determines discrete, coarse-grained versions of
the pointer observable $Z$ and the \me{}d \ob\ $E$;
 $$\align
Z^{\Cal R}:\, &i\mapsto Z_{i} 
\ :=\  Z(f^{-1}(X_i)),\ i\in\text{\bf I},\tag 3a\\
\vspace{2pt}
E^{\Cal R}:\, &i\mapsto E_i\ :=\ E(X_i),\ i\in\text{\bf I}.\tag 3b
\endalign
$$
The $Z^{\Cal R}$-value $i$ refers to the pointer reading
$f^{-1}(X_i)$ 
which, in turn, is correlated to the value set $X_i$ of the
meas\-ured observable $E$.
If $E$ itself is discrete there is a
natural (finest) reading scale $\Cal R$ such that 
$E=E^{\Cal R}$ and $Z^f$ = $Z^{\Cal R}$.
It should be noted that we have included the pointer function $f$ in the
definition of $Z^{\Cal R}$ so that the two discrete observables (3) do
have the same set of values.

\subheading{3.7}
We say that the (discrete) pointer observable $Z^{\Cal R}$ 
{\it has the value} $i$ in the state  $\pas(\vs)$ if
and only
the measurement outcome probability for this value equals one:
$\tr{\pas(\vs)Z_i} = 1$.
Since $\tr{TE_i} = \tr{\pas(\vs)Z_i}$, and, in general, $0\ne p^E_T(X_i)
\ne 1$, the pointer observable does not have a value at the end of the
measurement. It may, however, occur that the state $\pas(\vs)$
is a {\it mixture of eigenstates of} $Z^{\Cal R}$ {\it with the weights}
$p^E_T(X_i)$.
This is indeed
a necessary condition for the assertion that the pointer observable $Z$
has assumed a definite value with respect to a reading scale $\Cal R$ at the
end of the measurement $\Cal M$.
We go on to specify this case further.

\subheading{3.8}
We consider a measurement $\em$ of $E$ with a fixed reading scale $\Cal R$.
Any $X_i$, $i\in\bo I$, defines a (unnormalised) {\it conditioned state}:
$$
V_i(T) := I\otimes Z_i^{1/2}\,\vs\, I\otimes Z_i^{1/2},\tag 4
$$
the state of \sa\ on the condition that the pointer observable
$Z^{\Cal R}$ has value $i$.
The (trace) norm of this state is $\tr{V_i(T)} = \tr{\vs I\otimes Z_i}
= p^E_T(X_i)$, and the corresponding (normalised) reduced 
states, the  {\it  final component states} of 
\s\ and \a\  are:
 $$
\align
T_{\Cal S}(i,T)\ &:=\  p_T^E(X_i)^{-1}\,
\pa(V_i(T)),\tag 5a\\
T_{\Cal A}(i,T)\ &:=\  p_T^E(X_i)^{-1}\,
\pas(V_i(T)).\tag 5b\\
\endalign
$$
(If $p_T^E(X_i) = 0$, we put $T_{\Cal S}(i,T)=T_{\Cal A}(i,T) = O$).
The conditional
interpretation of the states (4) and (5) presupposes, however, 
that the pointer observable $Z^{\Cal R}$   has value $i$ in 
state $T_{\Cal A}(i,T)$,  that is, 
$\tr{T_{\Cal A}(i,T)Z_i} = 1$ 
for all $i$ and $T$ whenever $p^E_T(X_i)\ne 0.^5$ 
This requirement is always satisfied if the pointer observable is sharp.
In general this is a condition to be imposed
on the measurement scheme. We call it the
{\it pointer value-definiteness} condition and note that
it may be written in either of
the following equivalent forms:
 $$\align
&\tr{T_{\Cal A}(i,T)Z_i}\,=\,1\quad(\text{whenever }
p^E_T(X_i)\ne 0),\tag 6a\\
&Z_i\,T_{\Cal A}(i,T)\,=\,T_{\Cal A}(i,T),\tag 6b
\endalign
$$
for all $i\in\bo I$ and all initial states $T$ of \s. 

\subheading{3.9}
For any reading scale $\Cal R$ and any state
$T\in\sh$ one has 
$$
\pa(\vs)\ =\ \sum p_T^E(X_i)\,T_{\Cal S}(i,T);\tag 7
$$ 
this is to say that the final object state behaves additively with respect to
the pointer conditioning: that is,
the state of \s\ on the plain condition that the measurement has been
performed, is the same as the state of \s\ after the measurement
conditional on the fact
that the pointer value is registered with respect to the reading scale 
$\Cal R.^6$
Although it also holds true that for any  $i\in\bo I$ and $T\in\sh$
$$
T_{\Cal A}(i,T) = p_T^E(X_i)^{-1}Z_i^{1/2}\pas(\vs)Z_i^{1/2}, \tag 8
$$
it is not the case, in general, that the final apparatus state
$\pas(\vs)$ is conditioned with respect to $\Cal R$; thus, in general,
$$
\pas(\vs)\ \ne\ \sum p_T^E(X_i)\,T_{\Cal A}(i,T).\tag 9
$$
The requirement that $\pas(\vs)$ is a mixture
of the final component states $T_{\Cal A}(i,T)$ is therefore another
condition on the measurement${}^2$; we call it the {\it pointer
mixture condition}:
 $$
\pas\big(V(T\otimes T_{\Cal A})\big)\ =\ \sum p^E_T(X_i)\,T_{\Cal A}(i,T)\tag 10
$$
 for all initial states $T$ of \s.  

\subheading{3.10}
The pointer value-definiteness condition (6) and
the pointer mixture condition (10) imply that the final apparatus state is
a mixture of the pointer eigenstates $T_{\Cal A}(i,T)$ with the weights
$p_T^E(X_i)$; this means that the final apparatus state $\pas(\vs)$ is
conditioned with respect to the reading scale $\Cal R.^{2,6}$
One may consider the assumption that in addition to this,
the state $\pas(\vs)$ admits
the {\it ignorance interpretation} with respect to the decomposition (10):
that is, the apparatus [in state $\pas(\vs)$] is actually in one of the
component states $T_{\Cal A}(i,T)$,
and this is the case with the subjective probability $p^E_T(X_i)$.
As well known, such an interpretation of the mixed state $\pas(\vs)$
is extremely problematic and in most cases impossible; but if it were the case
then the pointer could be claimed to have a definite value $i$ (with
respect to a reading scale $\Cal R$) after the
measurement with the subjective probability $p^E_T(X_i)$.
Setting aside  the difficulties with the ignorance
interpretation (and thus with explaining the occurrence of
definite measurement outcomes in quantum mechanics), it still is
important to investigate more closely
the conditions (6) and (10) and to see how these possible properties of
a measurement are related to the structure of the final
state of the object system. 

\proclaim {{\bf 3.11} Theorem } Let $\Cal M$ be a
meas\-urement of an observable $E$ and $\Cal R$ any
reading scale. For any initial state $T$ of the object
system, the condition $a)$ implies the conditions $b)$ and $c)$:
 $$\align
 a)&\quad T_{\Cal S}(i,T)\cdot T_{\Cal S}(j,T)\,
\,=O\quad \text{for }i\neq j;\\
\vspace{3pt}
b)&\quad \Cal R_{\Cal A}\big(V(T\otimes T_{\Cal A})\big)
\ =\ {\textstyle{\sum}}
p^E_T(X_i)\, T_{\Cal A}(i,T)\quad \text{for all }i;\\
\vspace{3pt}
c)&\quad Z_i T_{\Cal A}(i,T)=T_{\Cal A}(i,T)\quad
 \text{for all }i.\\
 \endalign 
$$
If \m\ is a unitary \mt\ $\Cal M_U$, 
then $a)$ and $b)$ are equivalent conditions for any initial
vector state $T=\pee\fii$ of \s.
 \endproclaim 

 \demo{Proof} \newline
$a)\Rightarrow b)\& c)$:
 For each $i$, let $F_i$ be the support
projection of $T_{\Cal S}(i,T)$, that is, the smallest
projection $Q$ such that $Q T_{\Cal S}(i,T)=T_{\Cal S}
(i,T)$. Then one gets (for $i\ne j$):
$$
\align
T_{\Cal S}(i,T)\cdot T_{\Cal S}(j,T)\ =\ O\ 
 \ &\Leftrightarrow\ F_iT_{\Cal S}(j,T)=O\\
&\Leftrightarrow\ \tr{F_i\otimes I\, V_j(T)}=0\\
&\Leftrightarrow\ F_i\otimes Z_j^{1/2}\,
                  V(T\otimes T_{\Cal A})\,I\otimes Z_j^{1/2}  = O\\
&\Leftrightarrow\ F_i\otimes Z_j^{1/2}\,
                            V(T\otimes T_{\Cal A})^{1/2}=O\\
&\Rightarrow\ F_i\otimes Z_j\,
                            V(T\otimes T_{\Cal A})^{1/2}=O \tag{{$\alpha$}}
\endalign
$$
By the definition of $F_i$ one also has
$$
\align
&F_iT_{\Cal S}(i,T)=T_{\Cal S}(i,T)\\
&\Leftrightarrow\ F_i\otimes Z_i^{1/2}\,V(T\otimes T_{\Cal A})\,
I\otimes Z_i^{1/2}
 =I\otimes Z_i^{1/2}\,V(T\otimes T_{\Cal A})I\otimes Z_i^{1/2}\\
&\Rightarrow\ F_i\otimes Z_i\,V(T\otimes T_{\Cal A})^{1/2}
     =I\otimes Z_i\,V(T\otimes T_{\Cal A})^{1/2}
\tag{{$\beta$}}
\endalign
$$
Combining ($\alpha$) and ($\beta$) and using the fact that
$\sum Z_i = I$ yields
$$
 F_i\otimes I\,V(T\otimes T_{\Cal A})^{1/2}
     = F_i\otimes Z_i\,V(T\otimes T_{\Cal A})^{1/2} =
 I\otimes Z_i\,V(T\otimes T_{\Cal A})^{1/2}.\tag{{$\gamma$}}
$$
From this one obtains
$I\otimes Z_iV(T\otimes T_{\Cal A}) = 
I\otimes Z_i^2V(T\otimes T_{\Cal A})$,
which gives $c)$.
Using $c)$, one shows similarly  that 
$$
I\otimes Z_iV(T\otimes T_{\Cal A})^{1/2} = 
I\otimes Z_i^{1/2}V(T\otimes T_{\Cal A})^{1/2}.
$$
Inserting this in ($\gamma$), multiplying each term with its adjoint and summing
over $i$, one obtains
$$
\sum F_i\otimes I\,V(T\otimes T_{\Cal A})\,F_i\otimes I\ =\
\sum I\otimes Z_i^{1/2}\,V(T\otimes T_{\Cal A})\,I\otimes Z_i^{1/2}.
$$
Taking the partial trace with respect to $\hi$ 
finally yields $b)$.
\newline
 $b)\Rightarrow a)$: This implication will be shown for a
unitary \mt\ $\Cal M_U$ and for vector state preparations $T=\pee\fii$.
In that case $T_{\Cal A}=\pee\phi$ and $V(T\otimes
T_{\Cal A})=\pee{U(\fii\otimes\phi)}$. Denoting the
biorthogonal decomposition of this state as
$U(\fii\otimes\phi)=\sum_{nk} c_n\fii_{nk}\otimes\phi_{nk}$,
with $c_n>0$, we obtain $T_{\Cal A}(\Omega,T)=
\sum_{nk}|c_n|^2\pee{\phi_{nk}}$. Now $b)$ implies that all
the projections $Z_i$ commute with $T_{\Cal A}(\Omega,T)$. Therefore one can
choose the orthonormal system $\{\phi_{nk}\}$ such that
$Z_i\phi_{nk}=\phi_{nk}$ or $Z_i\phi_{nk}=0$. Thus there is
a renumbering of this system,
$\{\phi_{nk}\}=\{\phi_{i\ell}\}$, such that
$Z_i\phi_{i\ell}=\phi_{i\ell}$. It follows that there are
corresponding renumberings $\{\fii_{i\ell}\}=\{\fii_{nk}\}$
and $\{d_{i\ell}\}=\{c_n\}$ 
such that $U(\fii\otimes\phi)=\sum d_{i\ell}\fii_{i\ell}\otimes
\phi_{i\ell}$. Then $T_{\Cal S}(i,T)=\sum_\ell |d_{i\ell}|^2
\,\pee{\fii_{i\ell}}$. Since the subsets of vectors
$\fii_{i\ell}$ with different values of $i$ are mutually
disjoint and therefore orthogonal, one concludes that $a)$ holds.
This completes the proof.
\enddemo

\noindent
It can be demonstrated by means of examples that
the implication $b)\Rightarrow
a)$ need not hold  if the \mt\ is not unitary or
if the initial pointer state  is not pure$.^1$

\subheading{C. First kind and repeatable measurements}

\subheading{3.12}
A measurement $\Cal M$ of an observable $E$ is of the {\it first kind}
if the probability for a given result is the same both before and after 
the measurement, that is, for any $T\in\sh$
and for all $X\in\Cal F$,
$$
p^E_T(X) = p^E_{\pa(\vs)}(X).\tag 11
$$
Unitary measurement schemes with a coupling 
$U = e^{i\lambda A\otimes B}$, $\lambda\in\bo R$, $A$ (on $\hi$) 
and $B$ (on $\ha$)
self-adjoint, do  give rise to such measurements; we refer to
Sec.\ 8 for an analysis of this model.

\subheading{3.13}
A measurement $\Cal M$ of an observable $E$ is {\it repeatable} if its
repetition does not lead to a new result. One way to express the requirement
is the following: for any $T\in\sh$ and $X\in\Cal F$, if
$p^E_T(X) \ne 0$, then
$$
p^E_{T_{\Cal S}(X,T)}(X) = 1,\tag 12
$$
(where $T_{\Cal S}(X,T)$ is defined  by (3a), (4) and (5a) with $X=X_i$). 
Equivalently, $\Cal M$ is a repeatable $E$-measurement
if for any $T\in\sh$ and $X\in\Cal F$, for which
$p^E_T(X) \ne 0$, it holds true that
$$
E(X)\,T_{\Cal S}(X,T)\, =\, T_{\Cal S}(X,T).\tag 13
$$
Another basic result of  measurement theory is that
an observable $E$ which admits a repeatable measurement
is discrete$.^{4,7}$

\subheading{3.14}
According to (13),
a repeatable \mt\ drives the object system 
into an eigenstate of the \me{}d \ob\ $E:i\mapsto E_i$. 
The orthogonality conditions of Theorem 3.11
are then satisfied and the final apparatus state $\pas(\vs)$ is the mixture
of the eigenstates $T_{\Cal A}(i,T)$ of $Z:i\mapsto Z_i$ with the weights
$\tr{TE_i}$.

\subheading{3.15}
It is evident that repeatable measurements are also of the first kind.
However, as will be demonstrated in Sec.\ 8, a first kind measurement need
not be repeatable, though for sharp observables the two notions 
coincide$.^8$

\subheading{IV. Statistical dependence and correlations} 
 A meas\-urement \m\ of an observable $E$ brings the
compound object-apparatus system into an entangled state
$V(T\otimes T_{\Cal A})$. The possibility of transferring
information from \a\ to \s\ rests on the fact that this
state entails {\it statistical dependencies} between quantities
pertaining to these systems. Accordingly, three types of
{\it correlations} inherent in the state $V(T\otimes T_{\Cal A})$
are of special interest for characterising the \mt:
 $i$) correlations between the meas\-ured observable and the
pointer observable;
 $ii$) correlations between the corresponding values of
these observables; and
 $iii$) correlations between the final component states of
the two subsystems.
For their study it
is helpful to recall some basic notions and facts
concerning the relation between statistical dependence and
correlation.

\subheading{4.1}
Let $\mu$ be a probability measure on the
real Borel space $\big(\Rea^2,\Cal B(\Rea^2)\big)$,
and let $\mu_1$ and $\mu_2$ be the marginal meas\-ures of
$\mu$ with respect to a  Cartesian coordinate
system: for $X,Y\in\Cal B(\Rea)$,
$$
\mu_1(X) = \mu (X\times\Rea ), \ \ \
\mu_2(Y)=\mu(\Rea\times Y). \tag 14
$$ 
These marginal meas\-ures correspond to the coordinate projections
(random variables) 
$\pi_1:(x,y)\mapsto x$ and $\pi_2:(x,y)\mapsto y$ in the sense that
$\mu_i = \mu^{\pi_i}$, that is,
$\mu_i(X) = \mu^{\pi_i}(X) := \mu(\pi_i^{-1}(X))$ for all $X\in\Cal B(\Rea)$,
$i=1,2$.  
Assume
that the expectations and the variances of $\mu_i$
are well defined and finite: $\epsilon_i = \int
xd\mu_i(x)$, $\sigma_i^2 = \int (x-\epsilon_i)^2 d\mu_i(x)$, and let
$\epsilon_{12} = \int xy d\mu(x,y)$. The (normalised) {\it
correlation} of 
the marginal measures $\mu_1$ and $\mu_2$
in $\mu$ is then defined as:
 $$
\rho(\mu_1,\mu_2;\mu)\ 
:= \int \frac{(x-\epsilon_1)(y-\epsilon_2)}
{\sigma_1\,\sigma_2}d\mu (x,y)\ =\ 
\frac{\epsilon_{12}-\epsilon_1\epsilon_2}
{\sigma_1\,\sigma_2}\tag 15
$$
 (whenever $\sigma_1\neq 0\neq\sigma_2)$. The Schwarz
inequality entails $|\rho(\mu_1,\mu_2;\mu )|\leq
1$. The marginals $\mu_1,\mu_2$ are {\it
uncorrelated} if $\rho(\mu_1,\mu_2;\mu ) = 0$
(that is, $\epsilon_{12} = \epsilon_1\epsilon_2$),
{\it strongly correlated} if $\rho(\mu_1,\mu_2;\mu ) = 1$ 
(that is, $\epsilon_{12}-\epsilon_1\epsilon_2 =\sigma_1\sigma_2$),
and {\it strongly anticorrelated} if  $\rho(\mu_1,\mu_2;\mu ) = -1$
(that is, $\epsilon_{12}-\epsilon_1\epsilon_{2} =-\sigma_1\sigma_2$).
The strong correlation conditions can also be written in terms of 
the coordinate projections $\pi_1$ and $\pi_2$:
 $$
\align
\rho(\pi_1,\pi_2;\mu)=+1\ \ &\text{ iff }\ \
\pi_1\,=\,\frac{\sigma_1}{\sigma_2}(\pi_2-
\epsilon_2)+\epsilon_1\,=:\,\ell_+\circ\pi_2
\ \ \text{($\mu-$a.e.)},\tag 16a\\
\rho(\pi_1,\pi_2;\mu)=-1\ \  &\text{ iff }\ \
\pi_1\,=\,-\frac{\sigma_1}{\sigma_2}(\pi_2-
\epsilon_2)+\epsilon_1 \,=:\,\ell_-\circ\pi_2
\quad\text{($\mu-$a.e.)}.\tag 16b
\endalign
$$
(Here we have introduced the function
$\ell_\pm:y\mapsto\ell_\pm(y) := \pm\frac{\sigma_1}{\sigma_2}(y-\epsilon_2)
+\epsilon_1)$.
A case of special interest arises when the marginals $\mu_1$
and $\mu_2$ have the same (finite) first and second moments
so that $\epsilon_1 = \epsilon_2$, $\sigma_1=\sigma_2$. 
Then  one has:
 $$
\align
\rho(\mu_1,\mu_2;\mu) = +1\ &\text{ iff }\  
\epsilon_{12} = \epsilon_1^2 + \sigma_1^2,\tag 17a\\
&\text{ iff }\ \pi_1 = \pi_2\quad (\mu-a.e.),\\
\vspace{3pt}
\rho(\mu_1,\mu_2;\mu) = -1\ &\text{ iff }\  
\epsilon_{12} = \epsilon_1^2 -\sigma_1^2, \tag 17b\\
&\text{ iff }\ \pi_1 = -\pi_2 + 2\epsilon_1\quad (\mu-a.e.).
\endalign
$$

\subheading{4.2}
The notion of correlation can be applied to quantify the
degree of mutual  dependence of the marginal measures.
In order to avoid dealing with unnecessary
complications, we assume that $\mu_1$ and $\mu_2$ are no
$\{0,1\}$-valued \me{}s; equivalently, we let $\sigma_1\ne
0\ne\sigma_2$. $\mu_1$ and $\mu_2$ 
are {\it independent} if $\mu=\mu_1\times\mu_2$. Otherwise,
$\mu_1,\mu_2$  are {\it dependent}.  They are
{\it completely dependent} if there
is a (measurable) function $h:\Rea\to\Rea$ such that $\mu (X \times Y)
=\mu_2\big(h^{-1}(X)\cap Y\big)$ for $X,Y\in\Cal B(\Rea)$.  
That is, the marginal \me\ $\mu_2$ suffices to
determine the whole \me\ $\mu$. The relation of
complete dependence is symmetric with respect to the two
marginals only if $h$ is bijective.
This  is the  case of concern here.

\subheading{4.3}
It is evident that the statistical independence of 
$\mu_1,\mu_2$ implies $\rho(\mu_1,\mu_2;\mu)=0$. However, the latter
condition is not sufficient to ensure their independence.
(For a counter example, see, for instance ref. 9).  
On the other hand, eqs.\ (16a,b) show that strong
(anti)correlation entails complete dependence, the dependence being 
given by the linear function $\ell_\pm$.
Indeed, the condition $\pi_1 = \ell_\pm\circ\pi_2$ ($\mu$-a.e.) implies that
$\mu(X\times Y) = 0$ for all $X$ and $Y$ for which
$\ell_\pm^{-1}(X)\cap Y =\emptyset$. 
Thus, in particular, for any $X$ and $Y$, and with $X'$ denoting the
complement of $X$ one has
$\mu(X'\times\ell_\pm^{-1}(X)\cap Y)$ = 0
= $\mu(X\times\ell_\pm^{-1}(X')\cap Y)$.
The additivity properties of
$\mu$ allow one then to verify that for all $X,Y$,
$\mu_2(\ell_\pm^{-1}(X)\cap Y) = \mu(X\times Y)$, that is, $\mu_1$ 
and $\mu_2$ are completely dependent with $\ell_\pm$.
By a direct computation one can confirm that the converse
implication holds true whenever the function $h$ is
linear. 
Therefore, we have:
$$
\align
\rho(\mu_1,\mu_2;\mu)\ =\ +1\ \ &\text{iff}\ \mu_1,\mu_2\
\text{are completely dependent}\\ 
&\text{with}\  h(y)\ =\  ay+b,\ a>0, \tag 18a\\
\rho(\mu_1,\mu_2;\mu) =\ -1\ \  &\text{iff}\ \mu_1,\mu_2\
\text{are completely dependent}\\
&\text{with}\ h(y)\ =\ ay+b,\ a<0.\tag 18b
\endalign
$$
In both cases the constants are $a=\pm\sigma_1/\sigma_2$, 
$b= \epsilon_1- a\epsilon_2$, so that $h=\ell_\pm$.

\subheading{V. Strong correlations between observables}

\subheading{5.1}
 According to the condition (2), in
an $E$-\mt\ the initial $E$-outcome distribution is
recovered from the final $Z$-outcome distribution.
In addition to this basic requirement, a meas\-urement may
also establish complete statistical dependence between the
meas\-ured observable and the pointer observable after the
meas\-urement; that is, the observables $E$ and $Z^f$ may 
become strongly correlated in the final
object-apparatus state $V(T\otimes T_{\Cal A})$.  In order
to avoid technical complications in the formulation of
this correlation, we assume that the value space of $E$ is
the real Borel space, $\of = \rb$. Then for any state
$T\in\Cal S(\Cal H)$ the map
 $$
\mu: X\times Y\,\mapsto\, \tr{V(T\otimes T_{\Cal A}) 
E(X)\otimes Z^f(Y)}
 \tag 19
$$
 extends to a probability measure on $\big(\Rea^2,\Cal
B(\Rea^2)\big).^{10}$
The marginal distributions are
 $$
\align
\mu_1:X&\mapsto \tr{\pa(\vs)E(X)},
\tag 20a\\
\vspace{3pt}
\mu_2:Y&\mapsto  \tr{\pas(\vs)Z^f(Y)} =
\tr{TE(Y)}.
\tag 20b
\endalign
$$
Denoting the correlation of $\mu_1$ and $\mu_2$ in $\mu$ 
as
$\rho\big(E,Z^f;V(T\otimes T_{\Cal A})\big)$, we
say that the meas\-urement $\Cal M$ of $E$ produces {\it
strong observable-(anti)correlat\-ion} in  state $T$ if
this number equals $1\, (-1)$.  According to (18), this
occurs exactly when the probability measures (20a,b) 
are completely
dependent, with the function $\ell_\pm$. 
In order to analyze  the statistical dependence of $\mu_1$ and $\mu_2$
we shall make use of the
concept of a state transformer (also known as an instrument)
associated with a measurement.

\subheading{5.2} Consider a measurement $\em$ of $E$. 
Any $X\in\Cal F$
defines a non\-normalised state
$$
V_X(T) := I\otimes Z^{1/2}\,(f^{-1}(X))\vs\,I\otimes 
Z^{1/2}\,(f^{-1}(X)), \tag 21
$$
the (trace) norm of which is $\tr{V_X(T)} = p^E_T(X)$. Taking the
partial trace of $V_X(T)$ over $\ha$ one gets the (nonnormalised)
reduced state of \s,
$$
\Cal I_X(T) := \pa(V_X(T)). \tag 22
$$
For any $X\in\Cal F$ and  $T\in\sh$, $\tr{\Cal I_X(T)} = \tr{TE(X)}$,
and $T\mapsto\Cal I_X(T)$ is a (contractive) state transformation.
The mapping $\Cal I:X\mapsto\Cal I_X$ has the measure property
$\tr{\Cal I_{\cup X_i}(T)} = \sum\tr{\Cal I_{X_i}(T)}$ for any disjoint
sequence 
$(X_i)\subset\Cal F$ and for all $T\in\sh$.
Moreover, $\tr{\Cal I_\Omega(T)} = 1$ for any $T$.
We call $\Cal I$ the {\it state transformer} induced by the measurement
$\Cal M$. It describes the object system's state changes
under the measurement, and it uniquely defines the measured observable
via the relation $\tr{\Cal I_X(T)} = \tr{TE(X)}$. We note also that
$p^E_T(X)\,T_{\Cal S}(X,T) = \Cal I_X(T)$, and, in particular,
$\pa(\vs) = \Cal I_{\Omega}(T)$.

\subheading{5.3}
The probability measure (19) can be written as
$$
\mu(X\times Y) =
\tr{\Cal I_Y(T)E(X)} = \tr{\Cal I_X(\Cal I_Y(T))},\tag 23
$$
and the second marginal is
$\mu_2(Y) =\tr{\Cal I_Y(T)}$. 
The strong (anti-)correlation then amounts to
$$
\tr{\Cal I_X(\Cal I_Y(T))} = \tr{\Cal I_{\ell_\pm^{-1}(X)\cap Y}(T)}.\tag 24
$$
A special case of complete dependence arises with
$\ell_+$ being the identity function:
 $$
\tr{\Cal I_X(\Cal I_Y(T))}\,=\,\tr{\Cal I_{X\cap Y}(T)}.
\tag 25
$$
This relation is easily seen to coincide with (12)$.^8$ Thus,
if valid for all states $T$,
(25) expresses the repeatability of the \mt, and we may conclude that 
any repeatable \mt\ leads to strong
observable-correlations. 
The repeatability condition (25) is not necessary for the
strong observable-correlation (24).

\subheading{5.4}
Condition (25) implies, in particular, the equality of the
marginal meas\-ures $\mu_1,\mu_2$ of Eqs.\ (20a,b):  for all $X$,
 $$ 
p^E_T(X)\ =\ p^E_{\pa\left(V(T\otimes 
T_{\Cal A})\right)}(X).\tag 26
$$
This is just the first-kind property
of the \mt. It may occur
that these marginal \me{}s coincide irrespectively of
whether (25) holds or not; in that case conditions (17a,b) give the
relevant characterisations of strong (anti)correlations.

\proclaim{{\bf 5.5} Theorem}
 Let $\Cal M$ be a measurement of an observable $E$, and
let $\Cal R$ be any reading scale. Then $a)$ implies $b)$,
where:
 $$\align
a)\quad &E(X_i)T_{\Cal S}(i,T) = T_{\Cal S}(i,T)\ 
\text{ for all }\ T\in\Cal S(\hi),\ X_i\in\Cal R;\\  
\vspace{3pt}
b)\quad&\sigma\big(p^{E^{\Cal R}}_
{\pa(V(T\otimes T_{\Cal A}))}\big)\ne 0\ \text{and}\ 
\rho\big(E^{\Cal R},Z^{\Cal R};
V(T\otimes T_{\Cal A})\big) = 1\\ 
\vspace{1pt}
\quad &\text{for all }\ T\in\Cal S(\hi)\ \text{ with }\ 
\sigma\big(p_T^{E^{\Cal R}}\big)\ne 0.
\endalign
$$
 If the reading scale $\Cal R$ is finite, then $a)$
and $b)$ are equivalent.
 \endproclaim

\demo{Proof} 
The eigenstate condition $a)$ is equivalent with the repeatability
condition (with respect to $\Cal R$). Therefore, 
if $a)$ holds, then also $b)$ is true.
It remains to show that
$b)$ implies $a)$ whenever $\Cal R$ is finite.  According to
(18a), the statement
$\rho\big(E^{\Cal R},Z^{\Cal R};V(T\otimes T_{\Cal A})\big) = 1$
is equivalent to the complete
dependence, $\mu(i,j)\ =\ \mu_2(j)\, \delta_{i,\ell_+(j)}$, with
a bijective linear mapping $i=\ell_+(j)=aj+b$, $a>0$, between
those values $i,j$ for which $\mu_2(j)\ne 0$ (and hence
$\mu_1(i) =\mu_2\big(\frac 1a(i-b)\big)\ne 0$). \newline
 {\it Case 1.} Let $T$ be such that $0\ne \tr{TE_i}\ne 1$
for all $i\in \bo I$. Then $\mu(i,j)$ correlates, via
$i=\ell_+(j)=aj+b$, {\it all} values $j\in\bo I$ with values
$i\in \bo I$. Since $\ell_+$ is onto and monotonically increasing,
$\ell_+(j)=j$.  But the complete dependence
condition, with $\ell_+(j)=j$, is nothing but Eq.\ (25) (with respect to
$\Cal R$), which is
equivalent to $a)$.
\newline
 {\it Case 2.} Let $T$ be any state such that $0\ne\tr{TE_k}
\ne 1$ holds exactly for all $k\in\bo I_1$, a proper
nonempty subset of $\bo I$. Take any $T'$ for which
$0\ne\tr{T'E_l}\ne 1$ exactly for all $l\in\bo I'_1$,
the complement of $\bo I_1$. Then the reasoning of
Case 1 applies to $\hat T:=\frac 12 T+\frac 12 T'$. Hence,
$E_i\Cal I_i(\hat T)=\Cal I_i(\hat T)$ for all $i\in\bo
I$. Inserting in this equation the relation $\Cal
I_k(T')=O$, which holds for $k\in \bo I_1$, it follows that
$E_i\Cal I_i(T)=\Cal I_i(T)$ for $i\in \bo I_1$.  But this
relation holds trivially also for $i\in\bo I'_1$
since in that case $\Cal I_i(T)=O$. This completes the
proof.
 \enddemo

\subheading{VI. Strong correlations between values}

\subheading{6.1}
 The observable $E^{\Cal R}$ meas\-ured by the
scheme $\Cal M$ with the reading scale $\Cal R$ is
discrete.  One may therefore ask to what degree the {\it
values} of this observable and the pointer observable
$Z^{\Cal R}$ become correlated in the \mt. To
answer this question requires studying the correlation
$\rho\big(E_i,Z_i;V(T\otimes T_{\Cal A})\big)$ 
of the $i$-th values of these observables in the final
object-apparatus state,  
that is, the correlation of quantities
$E_i\otimes I$ and $I\otimes Z_i$ in the state $V(T\otimes T_{\Cal A})$:
$$
\rho\big(E_i,Z_i;V(T\otimes T_{\Cal A})\big)
\ =\ \frac{\epsilon_{12}\,-\,\epsilon_1\epsilon_2}{\sigma_1\sigma_2}.
\tag 27
$$ 
The respective quantities are easily determined:
$$
\align
\epsilon_{12} &= \tr{\Cal I_i^2(T)},\tag 28a\\
\vspace{3pt}
\epsilon_1 &= \tr{\Cal I(\bold I)(T)E_i},\tag 28b\\
\vspace{3pt}
\epsilon_2 &= \tr{TE_i},\tag 28c\\
\vspace{3pt}
\sigma_1^2 &=  \tr{\Cal I(\bold I)(T)E_i^2}-
\tr{\Cal I(\bold I)(T)E_i}^2,\tag 28d\\
\vspace{3pt}
\sigma_2^2 &=  \tr{\pas(V(T\otimes T_{\Cal A}))Z_i^2}-
\tr{TE_i}^2.\tag 28e
\endalign
$$
 Strong correlation is then equivalent to
 $$
\epsilon_{12}\,-\,\epsilon_1\epsilon_2=\sigma_1\sigma_2\tag 29
$$
whenever the right-hand side is nonzero.

\subheading{6.2}
Assume that the final
component state $T_{\Cal S}(i,T)$ is a 1-eigenstate of $E_i$
(whenever $p^E_T(X_i)\ne 0$); then one obtains
$\epsilon_{12}$ = $\epsilon_1$ = $\epsilon_2$ for all $T$.
It follows that $\epsilon_{12}-\epsilon_1
\epsilon_2=\sigma_1^2\le \sigma_1\sigma_2$ and thus
$\sigma_1\le\sigma_2$. On the other
hand, the relation $\epsilon_1=\epsilon_2=\epsilon_{12}$ 
together with
$\sigma_2^2\le \epsilon_2-\epsilon_2^2=\epsilon_{12}-
\epsilon_1\epsilon_2=\sigma_1^2$ implies
$\sigma_2\le\sigma_1$.  Therefore the correlation
$\rho(E_i,Z_i;V(T\otimes T_{\Cal A}))$ equals 1 whenever $0\ne
p_T^E(X_i)\ne 1$.

Another interesting implication of the eigenstate condition
$\epsilon_{12}=\epsilon_2$ and the ensuing equality $\sigma_2=\epsilon_2-
\epsilon_2^2$ is the fact that the state 
$T_{\Cal A}(i,T)$
is a
1-eigenstate of $Z_i$. With these observations we have established
the following result.

 \proclaim{{\bf 6.3} Theorem} Let \m\ be a \mt\ of an \ob\
$E$ and let $\Cal R$ be any reading scale. Then for any
state $T$ of \s,  $a)$ implies $b)$ and $c)$:
 $$\align
a)\quad & E_iT_{\Cal S}(i,T)=T_{\Cal S}(i,T)
\quad\text{for each }i;\\
\vspace{3pt}
b)\quad & \sigma\big(E_i\otimes I;V(T\otimes T_{\Cal
A})\big)\ne 0\ \text{ and }\ 
\rho\big(E_i,Z_i;V(T\otimes T_{\Cal A})\big)=1\\
&\text{for each }i\text{ with } 0\ne p^E_T(X_i)\neq 1;\\ 
\vspace{3pt}
c)\quad &T_{\Cal A}(i,T) 
\ \text{ is a 1-eigenstate of }Z_i
\quad\text{for each }i\ \text{ with } \ p^E_T(X_i)\ne 0.
\endalign
$$
\endproclaim

\noindent
This result entails that
a repeatable \mt\ is a strong value-correlation \mt.
Moreover, a necessary condition for \m\ to be a repeatable
\mt\ is that the final component state $T_{\Cal A}(i,T)$
of \a\ is a 1-eigenstate of the pointer observable, that is,
\m\ must fulfil the pointer value-definiteness condition. We
recall that this last property and in addition the pointer
mixture property arise already as consequences of the
mutual orthogonality of the component states $T_{\Cal S}(i,T)$
of \s\ (Theorem 3.11).
The notion of a correlation between values suggests that the
observables in question do have definite values; yet it
turns out that strong value-correlation does not require
pointer value-definiteness, nor repeatability. Even the
combination of $b)$ and $c)$ does not require the property
$a)$ to hold, as can be demonstrated by simple examples$.^1$

\proclaim{{\bf 6.4} Theorem}
 Let $\Cal M$ be a meas\-urement of a sharp observable
$E$ and $\Cal R$  any reading scale. For any
initial state $T$ of \s, $a)$ is equivalent to $b)\& c)$:
 $$\align
a)\quad &E_iT_{\Cal S}(i,T)=T_{\Cal S}(i,T)
\quad\text{for each }i;\\
\vspace{3pt}
b)\quad & \sigma\big(E_i\otimes I;V(T\otimes T_{\Cal A}) 
\big)\ne 0\ \text{ and }\ 
\rho\big(E_i,Z_i;V(T\otimes T_{\Cal A})\big)=1\\
&\text{for each }i \text{ with } 0\ne p^E_T(X_i)\neq 1;\\
\vspace{3pt}
c)\quad &T_{\Cal A}(i,T)
\ \text{ is a 1-eigenstate of }Z_i
\quad\text{for each }i\ \text{ with } \ p^E_T(X_i)\ne 0.
\endalign
$$
\endproclaim 
 \demo{Proof} In view of Theorem 6.3 we only need to show
that $b)\& c)$ implies $a)$. Hence let
$\epsilon_{12}-\epsilon_1 \epsilon_2=\sigma_1\sigma_2$ hold
for each $i$.  
Condition $c)$ implies $\sigma_2^2=\epsilon_2-\epsilon_2^2$.
Similarly the relation $E_i^2=E_i$ implies $\sigma_1^2=
\epsilon_1-\epsilon_1^2$. 
From Eqs.\ (28) we obtain $\epsilon_{12}\le\epsilon_1$,
$\epsilon_{12}\le\epsilon_2$, and therefore
 $$
\sigma_1\sigma_2=\epsilon_{12}-\epsilon_1\epsilon_2 
\le \sigma_1^2,\quad
\sigma_1\sigma_2=\epsilon_{12}-\epsilon_1\epsilon_2 
\le \sigma_2^2.
$$
This implies $\sigma_1=\sigma_2$. On the other hand,
 $$
\epsilon_1\epsilon_2+\sigma_1\sigma_2=\epsilon_{12}
\le \epsilon_1=\sigma_1^2+\epsilon_1^2,\quad
\epsilon_1\epsilon_2+\sigma_1\sigma_2=\epsilon_{12}
\le \epsilon_2=\sigma_2^2+\epsilon_2^2.
$$
 Using $\sigma_1=\sigma_2$, one concludes that $\epsilon_1=
\epsilon_2=\epsilon_{12}$. But the last equation is
equivalent to $a)$. This completes the proof.
 \enddemo

\subheading{VII. Strong correlations between final 
component states}

\subheading{7.1}
 In the two preceding sections it was demonstrated in which
way strong observable and value correlations serve as
characterisations of repeatable \mt{}s. The corresponding
eigenstate condition $E_iT_{\Cal S}(i,T)=T_{\Cal S}(i,T)$
entails, in particular, that the final component states of
the object associated with different outcomes $i,j$ are
mutually orthogonal,
 $T_{\Cal S}(i,T)\cdot T_{\Cal S}(j,T)\ =\ 0$.
 In some cases this orthogonality can be characterised in
terms of strong correlations between the final component
states of \s\ and \a.

 Consider a measurement scheme $\Cal M$
of an observable $E$ with respect to a reading scale
$\Cal R$.  We say that $\Cal M$, with $\Cal R$, is a {\it strong
state-(anti)correlation} meas\-urement of $E$ if for each
initial state $T$ of $\Cal S$ it correlates strongly the
final component states $T_{\Cal S}(i,T)$ and $T_{\Cal
A}(i,T)$ of the object and the apparatus. This calls for the
study of the correlation $\rho\big(T_{\Cal S}(i,T),T_{\Cal
A}(i,T);V(T\otimes T_{\Cal A})\big)$ of the probability
meas\-ure defined by the self-adjoint operators 
$T_{\Cal S}(i,T)\otimes I$ and 
$I\otimes T_{\Cal A}(i,T)$ and the
final object-apparatus  state $V(T\otimes T_{\Cal A})$. 
 
\proclaim {{\bf 7.2} Theorem} Let $\Cal M$ be a
meas\-urement of an observable $E$ and $\Cal R$ any
reading scale. For any initial state $T$ of the object
system for which the component states $T_{\Cal S}(i,T)$ and
$T_{\Cal A}(i,T)$ are vector states, $a)$ is equivalent to $b)\& c)$:
 $$\align
 a)&\quad T_{\Cal S}(i,T)\cdot T_{\Cal S}(j,T)\,
\,=O\quad \text{for }i\neq j;\\
\vspace{3pt}
b)&\quad \rho\big(T_{\Cal S}(i,T),T_{\Cal
A}(i,T);V(T\otimes T_{\Cal A})\big)=1\quad\text{for each }i
\text{ with }\ 0\ne p^E_T(X_i)\neq 1;\\
\vspace{3pt}
c)&\quad T_{\Cal A}(i,T)
\quad\text{is a 1-eigenstate of}\ Z_i
\quad\text{for each }i \text{ with }\ 0\ne p^E_T(X_i)\neq 1.
\endalign
$$
 \endproclaim 
 \demo{Proof} 
The equivalence is shown to hold under  the 
assumptions $T_{\Cal S}(i,T)=\pee{\fii_i}$ and 
$T_{\Cal A}(i,T)=\pee{\phi_i}$. These
two relations imply that $I\otimes Z_i^{1/2}\,
V(T\otimes T_{\Cal A})\, I\otimes Z_i^{1/2}$ is a
vector state of the product form, that is,
 $$
 I\otimes Z_i^{1/2}\,V(T\otimes T_{\Cal A})\, I\otimes
Z_i^{1/2}= p_T^E(X_i)\,\pee{\fii_i\otimes\phi_i}.
\tag\text{$\alpha$}
$$
If $a)$ holds, then by
Theorem 3.11, \m\ fulfils the pointer  value-definiteness condition $c)$.
Thus for both implications one can make use of the fact that
$Z_i\phi_i=\phi_i$. Then ($\alpha$)
implies
 $$
I\otimes \pee{\phi_i}\,V(T\otimes T_{\Cal A})\,
I\otimes \pee{\phi_i}=p_T^E(X_i)\,\pee{\fii_i\otimes\phi_i}.
$$
With this one computes:
 $$\align
\epsilon_{12}&=\tr{\pee{\fii_i}\otimes\pee{\phi_i}\,
             V(T\otimes T_{\Cal A})}=p_T^E(X_i),\\
\epsilon_1&=\tr{\pee{\fii_i}\,
     \pa\big(V(T\otimes T_{\Cal A})\big)}=
{\textstyle{\sum_j}} 
p_T^E(X_j)\,\tr{\pee{\fii_i}\,\pee{\fii_j}},\\
\epsilon_2&=
\tr{I\otimes\pee{\phi_i}\,V(T\otimes T_{\Cal A})}=
p_T^E(X_i),\\
\sigma_1^2&=\epsilon_1-\epsilon_1^2,\\
\sigma_2^2&=\epsilon_2-\epsilon_2^2.
\endalign
$$
$a)\Rightarrow b)$: 
$a)$ is equivalent to
$\tr{\pee{\fii_i}\,\pee{\fii_j}}=\delta_{ij}$, one has
$\epsilon_1=\epsilon_2=\epsilon_{12}$, and $\sigma_1=\sigma_2$. 
Thus 
$\epsilon_{12}-\epsilon_1\epsilon_2=\sigma_1\sigma_2$, that
is, $b)$.\newline 
$b)\& c)\Rightarrow a)$: Let
$\epsilon_{12}-\epsilon_1\epsilon_2=\sigma_1\sigma_2$.
Using the inequalities $\epsilon_{12}-\epsilon_1\epsilon_2\le\sigma_k^2$,
$k=1,2$, one concludes that $\sigma_1=\sigma_2$. Since
$\epsilon_{12} =\epsilon_2$, one also has $\epsilon_{12}-
\epsilon_1\epsilon_2=\sigma_2^2=\epsilon_2-\epsilon_2^2$, and
therefore $\epsilon_1=\epsilon_2$. But from the definition
of $\epsilon_1$ one has $\epsilon_1\ge\epsilon_2$, so that
the equality of these numbers implies 
$\tr{\pee{\fii_i}\,\pee{\fii_j}}=0$ whenever $i\ne j$, that
is $a)$. \newline
This completes the proof.
\enddemo

\subheading{7.3}
One may also ask whether the requirement of strong
correlation between the final \s\ and \a\ states
$\pa\big(V(T\otimes T_{\Cal A}) \big)$ and
$\pas\big(V(T\otimes T_{\Cal A})\big)$ imposes any
constraint on the \mt\ scheme under consideration. That this
cannot be expected in general can be seen in the case of a
unitary \mt\ $\Cal M_U$. Note first that the
reduced states of $\pee\us$ have the same spectra, including
multiplicities. The spectral decompositions can be given in
terms of orthonormal systems $\{\fii_i\}$, $\{\phi_i\}$
defined by the biorthogonal decomposition
$\us=\sum_ic_i\fii_i\otimes\phi_i$ ($c_i>0$), and a
straightforward calculation shows that
 $$
\rho\Big(\pa\big(\pee\us\big),\pas\big(\pee\us\big);
\pee\us\Big)=1.\tag 30
$$
 Hence these states are always strongly correlated.

\subheading{VIII. Examples}

\subheading{8.1}
A particularly interesting class of measurements arises
if the coupling is generated by a unitary map of the form
$$
U  \ =\ e^{i\lambda A\otimes B},\tag 31
$$
where 
$A$ and $B$ are self-adjoint operators in $\hi$ and $\ha$, respectively,
and $\lambda\in\bo R$ is a coupling constant.
The operator $A$ is usually taken to represent the (sharp) observable one
aims to measure. In order to specify the full measurement scheme and thus the
actually measured observable, one neeeds to choose the
pointer observable $Z$ and fix the initial preparation $T_{\Cal A}$ of 
the apparatus; the measured observable is then given by eq.\ (2).
Using the spectral decomposition of $A$, $A = \int aE^A(da)$,
and denoting
$$
T_{\Cal A}^{\lambda a} := e^{i\lambda aB}T_{\Cal A}e^{-i\lambda aB},\tag 32
$$
the final apparatus state, for $T\in\Cal S(\hi)$, assumes the form
$$
\pas(UT\otimes T_{\Cal A}U^*) = \int\tr{TE^A(da)}\,T_{\Cal A}^{\lambda a}.\tag 33
$$
Since it is of interest to
compare the measured observable $E$ with $E^A$ we assume from the
outset that the value space of $Z$ is $\rb$.
In view of the coupling constant $\lambda$ ($\ne 0$) it is also convenient to
introduce a pointer function $f(x)=\lambda^{-1}x$.
The observable $E$ meas\-ured  by the scheme
$\vas\ha,Z,T_{\Cal A},f,U\oik$ takes then the following form:
for any $X\in\Cal B(\bo R)$,
$$
E(X) 
\ =\  \int_\Rea\, \tr{T_{\Cal A}^{\lambda a}Z(\lambda X)}\, E^A(da).
\tag 34
$$

The structure of the operators $E(X)$  show that in general the 
meas\-ured observable $E$ is not the sharp observable $E^A$, 
but a smeared version of it$.^{11}$ 
One may  ask which
choices of $Z$ and $T_{\Cal A}$ would possibly yield
$E = E^A$. 
Obviously, this
is the case if and only if for ($E^A$-almost) all $a\in\Rea$,
$\tr{T_{\Cal A}^{\lambda a}Z(\lambda X)} = \chi_{{}_X}(a)$, 
where $\chi_{{}_X}$ is the characteristic function of the set $X$.

The measurement scheme thus defined  
is always of the first kind:
the measurement outcome probabilities for $E$ are the same both before and 
after the measurement; for any $T\in\Cal S(\hi)$ and for all $X\in\Cal B(\bo R)$,
$$
\tr{TE(X)} = \tr{UT\otimes T_{\Cal A}U^*E(X)\otimes I}.\tag 35
$$
It may also be noticed that the measurement does neither alter
the measurement outcome probabilities of $E^A$, though, as a rule, it is not
a measurement of $E^A$. In fact, if the \mt\ were an $E^A$-measurement,
it would also be repeatable (3.15) and $A$ would thus have to be
discrete,
$A = \sum a_iE^A(\{a_i\})$ (3.13). In that case the measurement would also
produce all the strong correlations discussed in the previous sections. 
In general, this is, however, not the case.

Consider next this measurement scheme with a fixed reading scale $\Cal R$.
The pointer observable $Z$ as well as the measured observable $E$
become discretized,
$$
\align
&Z^{\Cal R}:\ i\mapsto Z_i := Z(\lambda X_i),\tag 36a\\
&E^{\Cal R}:\ i\mapsto E_i := E(X_i),\tag 36b
\endalign
$$
and  the final component states are 
$$
\align
T_{\Cal S}(i,T) &= p^E_T(X_i)^{-1}\pa(V_i(T))\\
&= p^E_T(X_i)^{-1}\,\int\int\,
\pa\bigl(E^A(da)TE^A(da')\otimes Z_i^{1/2}e^{iaB}T_{\Cal A}e^{-ia'B}
Z_i^{1/2}\bigr)
\\
&= p^E_T(X_i)^{-1}\,\sum t_n\sum\, L_{kn}^i\,T\,L_{kn}^i{}^*\tag 37a\\
&{}\quad\text{with}\ L_{kn}^i := \int\ip{\psi_k}{Z_i^{1/2}\phi^{\lambda a}_n}
\, E^A(da)\in\Cal L(\hi),\\
&{{}\qquad\ \ } T_{\Cal A}\, = \sum t_n\pee{\phi_n}\quad \text{(spectral
decomposition)}\\
&{{}\qquad\ \ } \phi_n^{\lambda a} = e^{i\lambda aB}\phi_n,\\
&{{}\qquad\ \ } \{\psi_k\}\subset \ha\ \text{an orthonormal basis},\\
T_{\Cal A}(i,T) &= p^E_T(X_i)^{-1}\pas(V_i(T))\\
&= p^E_T(X_i)^{-1}\,\int\, \tr{TE^A(da)}\, Z_i^{1/2}T_{\Cal A}^{\lambda a}Z_i^{1/2},
\tag 37b
\endalign
$$
(provided that $p^E_T(X_i)\ne 0$).
If $E_i^2 = E_i$ for all $i\in\bo I$, the  measurement is repeatable 
with respect to $\Cal R$, and
$$
\align
&E(X_i)T_{\Cal S}(i,T) = T_{\Cal S}(i,T),\tag 38a\\
&T_{\Cal S}(i,T)\cdot T_{\Cal S}(j,T) = O, \ i\ne j, \tag 38b
\endalign
$$
in which case the implications of theorems 3.11, 5.5, 6.3, 6.4, and 7.2 all
hold true. We specify next two instances of the above model, one in which
$E_i^2 = E_i$ and another one with $E_i^2 < E_i$.

\subheading{8.2}
Consider a discrete observable $A = \sum a_kE^A(\{a_k\})$, and assume that the
set of eigenvalues of $A$ is closed.
As the apparatus (or a part of it, called probe) take a particle moving
in one-dimensional space, so that $\ha=L^2(\Rea)$,
and couple $A$ with its momentum $P_{\Cal A}$
according to (31).
Since the momentum generates translations on the position, it is
natural to choose the position 
$Q_{\Cal A}$ conjugate to $P_{\Cal A}$ as the pointer observable. 
Assuming that the initial state of \a\ is a vector state $\pee{\phi}$, then,
in the position representation (for \a) one has $\phi^{\lambda a_k}(x)=
\phi(x+\lambda a_k)$, with 
$\phi^{\lambda a_k} = e^{i\lambda a_kP_{\Cal A}}\phi$. 
Assuming that the spacing between
the eigenvalues 
$a_k$ is greater than $\frac\delta\lambda$ and that $\phi$ is supported in 
$\bigl(-\frac\delta{2},\frac\delta{2}\bigr)$, then the 
pointer states $\phi^{\lambda a_k}$ are
supported in the mutually disjoint sets 
$\lambda I_k$, where
$I_k =\bigl(a_k-\frac\delta{2\lambda},a_k +\frac\delta{2\lambda}\bigr)$. 
Introducing yet another pointer function $g$ such that
$g(I_k) = \{a_k\}$ for each $k$, and $g\big((\cup_kI_k)'\big) \subset
\{a_k:k=1,2,\cdots\}'$, one obtains from eq.\ (34)
$$
E\big(\{a_k\}\big) \ = \
\sum \ip{\phi^{\lambda a_i}}{E^{Q_{\Cal A}}(\lambda I_k)
\phi^{\lambda a_i}}\,E^A(\{a_i\})\ =\ E^A(\{a_k\}),
\tag 39
$$
for each $k$,
which shows that the observable meas\-ured by this scheme is indeed $E^A$. 
The measurement is repeatable, even a L\"uders measurement with the
state transformer $T\mapsto\Cal I_k(T) = \pa(V_k(T)) =
E^A(\{a_k\})TE^A(\{a_k\})$, 
and all the correlations introduced above are strong.

As an elementary quantum optical application, one may consider the
measurement of the number observable $N = a^*a$ of a (single-mode) signal
field by means of coupling it, via $e^{i\lambda N\otimes b^p}$, with one
of the quadrature components
$b^p = \frac i{\sqrt 2}(b^*-b)$, say,
of another single-mode (probe) field,
and using the other quadrature component $b^q= \frac 1{\sqrt 2}(b^*+b)$
as the readout observable.
With the above choices of the initial probe state $\phi$
and the pointer functions one obtains a number measurement.
It may be noted that neither the beam splitter coupling nor the number-number
coupling leads to a sharp number measurement$.^{11}$

\subheading{8.3}
The second illustration of the above model concerns the case of $A$
being a continuous observable, such as the position of a particle or
a quadrature component of a single-mode electromagnetic field.
Using the quantum optical nomenclature, we take 
$A =a^q=\frac 1{\sqrt 2}(a^*+a)$, 
the amplitude quadrature of the (single-mode) signal field with
the bosonic annihilation and creation operators $a,a^*$. 
For $B$ we take the corresponding quadrature component $b^q$ 
of  a (single-mode) probe field,
with the annihilation and creation operators $b,b^*$.
Using the phase quadrature 
$b^p := \frac i{\sqrt 2}(b^*-b) = \int_\Rea xZ(dx)$ of the probe field
as the readout observable, 
and assuming that the probe field is prepared in a vector state $\pee\phi$
determines the measured observable (34) to be of the form:
$$
\eqalign{
E(X) \ 
&= \int\int|\hat{\phi}|^2(y-\lambda x)\, \chi_{{}_{\lambda X}}(y)dy\,E^A(dx)\cr
&= \int|\hat{\phi}|^2(y-\lambda a^q)\, \chi_{{}_{\lambda X}}(y)dy\cr
&\equiv\ (e_{_\lambda}*\chi_{{}_{ X}})( a^q),\cr
e_{_\lambda}(y)\ &:=\ \lambda|\hat{\phi}|^2(-\lambda y),
}\tag 40
$$
where
$e_{_\lambda}*\chi_{{}_{ X}}$
denotes the convolution of the density function $e_{_\lambda}$  with
the characteristic function  of the  set $X$,
and $\hat{\phi}$ is the Fourier transform of $\phi$.

In the present case the measured observable is the \pov\
$E:X\mapsto (e_{_\lambda}*\chi_{{}_{X}})(a^q)$
and not  the spectral measure $X\mapsto \chi_{{}_X}(a^q)$ of $a^q$; 
this is to say that the 
measured field observable is not the amplitude quadrature $a^q$
but a smearing of it.
In fact, if $e$ were replaced by a delta function (concentrated at 0), 
then (40) would simply give the amplitude quadrature $a^q$. 
But this can never occur
since the readout observable  $b^p$ has no eigenstates, that is,
the initial state of the probe field cannot be so chosen that $e$ were
a delta function.
We observe also that the measurement is not repeatable
(since $E$ is not discrete) though still
of the first kind. Therefore, the strong correlations
are not guaranteed from the outset but need to be studied separately.

Before calculating the observable-correlation produced by the measurement
we compare the variance of $E$ with that of $a^q$ 
in a vector state $\pee\fii$.
Direct application of eq.\  (2) yields
(assuming that the involved quantities are finite)
$$
\text{Var}\,(E,\fii)\ =\ 
\text{Var}\,(a^q,\fii)
+ \frac 1{\lambda^2}\text{Var}\,(b^p,\phi). \tag 41
$$
The initial state $\pee\phi$ of the probe field can be
chosen  such that $\ip{\phi}{b^p\phi} = 0$. 
In this case the measured observable appears, 
in view of the first moments, as the amplitude quadrature $a^q$. 
However, the second moment 
$\ip{\phi}{(b^p)^2\phi}$  never equals 0, 
meaning that 
$\text{Var}\,(E,\pee\fii)$ is strictly greater than 
$\text{Var}\,(a^q,\pee\fii)$. 
However, in the limit of strong coupling, $\lambda\to\infty$, the
\mt\ noise term
$\frac 1{\lambda^2}\text{Var}\,(b^p,\pee\phi)$ tends to zero.
In any case, this
shows once more that the actually measured observable is not the
amplitude quadrature.

The observable-correlation  produced by the measurement
is now found to be
$$
\rho(E,Z^f;\pee{\us}) = \frac {\text{Var}\,(a^q,\pee\fii)}{\text{Var}\,
(E,\pee\fii)},\tag 42
$$
a quantity always strictly less than 1. The measurement,
though of the first kind,
does never lead to strong observable-correlation.
Yet,
$$
\lim_{\lambda\to\infty}\,\rho(E,Z^f;\pee{\us}) = 1.\tag 43
$$

In order to discuss the value- and state-correlations produced by the 
measurement scheme one needs to introduce a reading scale $\Cal R$.
The discrete observable $E^{\Cal R}:i\mapsto E_i$
thus measured is 
$$
E_i = (e_{_\lambda}*\chi_{{}_{ X_i}})(a^q),\tag 44
$$
whereas the final component states (37a) are of the form:
$$
\align
&T_{\Cal S}(i,\pee\fii) = \ip{\fii}{E_i\fii}^{-1}\int_{\lambda X_i}\,
L_y\pee\fii L_y^*\,dy \tag 45\\
&\text{with}\ L_y := \hat\phi(y-\lambda a^q).
\endalign
$$
Neither the eigenvalue condition (38a) nor the orthogonality condition
(38b) can be satisfied for all initial vector states of the signal 
field. Therefore the strong value and state-correlations
cannot be inferred by using theorems 6.3 and 7.2.
Still the value-correlation is always strong:
$\rho(E_i,Z_i;\pee{\us}) = 1$ for all $i$ and for any $\pee\fii$
for which $\ip{\fii}{E_i\fii}\ne 0$. Indeed, due to the commutativity of
the operators $L_y$ of eq.\ (45) with $E_i$, $\epsilon_{12}$ of eq.\ (28a)
equals $\ip{\fii}{E_i^2\fii}$; furthermore the first kind property of
the measurement and the sharp pointer yield for (28b-d): 
$\epsilon_1 = \epsilon_2 = \ip{\fii}{E_i\fii}$, and
$\sigma_1^2 = \sigma_2^2 = \text{Var}\,(E_i,\pee\fii)$. 
Therefore $\epsilon_{12}-\epsilon_1\epsilon_2=\sigma_1\sigma_2$, so that
$\rho(E_i,Z_i;\pee{\us}) =
\text{Var}\,(E_i,\pee\fii)/ \text{Var}\,(E_i,\pee\fii) = 1$.
Finally,
a direct computation of the state-correlation 
$\rho(T_{\Cal S}(i,\pee\fii), T_{\Cal A}(i,\pee\fii);\pee{\us})$
shows that this number is not, in general, equal to one.

\subheading{IX. Conclusion}

\noindent
In this paper we have investigated possible properties of the final
component states of the object system and the apparatus (or probe)
arrived at in a quantum \mt, properties which must be required
if the occurrence of definite \mt\ outcomes is to be understood
as the conjunction of pointer value definiteness ({\gr pvd}), pointer mixture
property ({\gr pm}), plus the ignorance interpretation for the final reduced
apparatus state. According to Theorem 3.11, the properties ({\gr pvd)}
and ({\gr pm}) are ensured if the final component states of the object
system are mutually orthogonal.
Considering initial states of \s\ which are vector states,
this latter condition is also necessary
for ({\gr pm}) in the case of a unitary \mt\ $\Cal M_U$, where ({\gr
pvd}) is automatically fulfilled since the pointer is a sharp \ob.
The orthogonality of the states $T_{\Cal S}(i,T)$ is not always
guaranteed$.^{12}$

Next we have considered conditions for strong correlations between
observables, their values, or between the final component states of
\s\ and \a. It turns out that repeatable \mt{}s give strong observable-
as well as strong value-correlations (Theorems 5.5, 6.3, 6.4).
Furthermore, strong observable-correlation for finite reading scales
entails repeatability and thus the orthogonality of the states
$T_{\Cal S}(i,T)$ and hence ({\gr pvd}), via 3.11. On the other hand,
strong value-correlation may occur independently of ({\gr pvd}).
Finally, strong state-correlation may occur under more general
circumstances than the other correlations since it is independent
of the repeatability property, but its implying the orthogonality
of the final component states of \s\ may be limited to the case
where these states are vector states. However in that case, and for a
unitary measurement $\Cal M_U$, strong
state-correlation is  equivalent to the said orthogonality and thus to
the pointer mixture condition. 

In conclusion, we wish to emphasize that
our investigation provides an illustration of how interpretational
demands entail formal constraints on \mt{}s that may or may not
be fulfilled in a concrete case. These formal features have thus to be made
explicit if the consistency of an interpretation is to be demonstrated.
With these findings we believe to have settled the questions left open in
previous work$.^{2}$

\subheading{Acknowledgements}
Part of this work was carried out while one author (PB) was
Visiting Research Scholar at the Lyman Laboratory of Physics, Harvard
University, Cambridge, MA. This visit was funded by means of a
Feodor Lynen Fellowship extended to him by
the Alexander von Humboldt-Foundation, Bonn. Support and hospitality
of Harvard University are gratefully acknowledged.

% }
%\endgroup

\vskip 3pt
\subheading{References}
\vskip 3pt

{\parindent=20pt\parskip= 2pt

\item{1.\ } P. Busch, P. Lahti,  P. Mittelstaedt, {\it The Quantum Theory
of Measurement}, {\bf LNP m2} (Springer-Verlag, Berlin, 1991).
Second revised edition forthcoming.

\item{2.\ }  G. Cassinelli, P. Lahti, {\it Nuovo Cimento B}  
{\bf 108}, 45 (1993).

\item{3.\ } D. F. Walls, G. J. Milburn,
{\it Quantum Optics} (Springer-Verlag, Berlin, 1994).

\item{4.\ } M. Ozawa, {\it J. Math. Phys.} {\bf 25}, 79 (1984).

\item{5.\ }  G. Cassinelli, N. Zanghi, {\it Nuovo Cimento B}  
{\bf 73}, 237 (1983).

\item{6.\ }  G. Cassinelli, N. Zanghi, {\it Nuovo Cimento B}  
{\bf 79}, 141 (1984).

\item{7.\ }  A. \L uczak,  Instruments on von Neumann algebras,
Institute of Mathematics, \L\'od\'z University, Poland, 1986.

\item{8.\ } P. Lahti, P. Busch, P. Mittelstaedt, {\it J. Math. Phys.}
{\bf 34}, 2770 (1991).

\item{9.\ } P. Halmos, {\it Measure Theory}, {\bf GTM 18}
(Springer-Verlag, New York, 1988).

\item{10.\ } S. K. Berberian {\it Notes on Spectral Theory},
(D. Van Nostrand Company, Inc. Princeton, 1966).

\item{11.\ } P. Busch, M. Grabowski, P. Lahti,
{\it Operational Quantum Physics}, {\bf LNP m31} (Springer-Verlag, Berlin,
1995).

\item{12.\ } E. Beltrametti, G. Cassinelli, P. Lahti,
{\it J. Math. Phys.} {\bf 31} (1990) 91.

\bye